\documentclass[twocolumn,trackchanges]{aastex631}
\usepackage{float}

\def\arcsec{\hbox{\ensuremath{^{\prime\prime}}}}
\def\farcs{\hbox{\ensuremath{.\!\!^{\prime\prime}}}}

\newcommand{\mbh}{\ensuremath{M_\mathrm{BH}}}

\newcommand{\rg}{\ensuremath{r_g}}
\def\farcs{\hbox{\ensuremath{.\!\!^{\prime\prime}}}}

\graphicspath{{./}{figures/}}
\begin{document}

\title{Black Hole Mass Measurements of Early-Type Galaxies NGC 1380 and NGC 6861 Through ALMA and HST Observations and Gas-Dynamical Modeling\footnote{Based on observations made with the NASA/ESA Hubble Space Telescope, obtained at the Space Telescope Science Institute, which is operated by the Association of Universities for Research in Astronomy, Inc., under NASA contract NAS5-26555. These observations are associated with programs 15226 and 11712.}}

\author[0000-0003-2632-8875]{Kyle M. Kabasares}
\affiliation{Department of Physics and Astronomy, 4129 Frederick Reines Hall, University of California, Irvine, CA, 92697-4575, USA}

\author[0000-0002-3026-0562]{Aaron J. Barth}
\affiliation{Department of Physics and Astronomy, 4129 Frederick Reines Hall, University of California, Irvine, CA, 92697-4575, USA}

\author[0000-0002-3202-9487]{David A. Buote}
\affiliation{Department of Physics and Astronomy, 4129 Frederick Reines Hall, University of California, Irvine, CA, 92697-4575, USA}

\author[0000-0001-6301-570X]{Benjamin D. Boizelle}
\affiliation{George P. and Cynthia Woods Mitchell Institute for Fundamental Physics and Astronomy, 4242 TAMU, Texas A\&M University, College Station, TX, 77843-4242,}
\affil{Department of Physics and Astronomy, 284 ESC, Brigham Young University, Provo, UT, 84602, USA}

\author[0000-0002-1881-5908]{Jonelle L. Walsh}
\affiliation{George P. and Cynthia Woods Mitchell Institute for Fundamental Physics and Astronomy, 4242 TAMU, Texas A\&M University, College Station, TX, 77843-4242,}

\author[0000-0002-7892-396X]{Andrew J. Baker}
\affiliation{Department of Physics and Astronomy, Rutgers, the State University of New Jersey, 136 Frelinghuysen Road, Piscataway, NJ 08854-8019, USA}
\affiliation{Department of Physics and Astronomy, University of the Western Cape, Robert Sobukwe Road, Bellville 7535, South Africa}

\author[0000-0003-2511-2060]{Jeremy Darling}
\affiliation{Center for Astrophysics and Space Astronomy, Department of Astrophysical and Planetary Sciences, University of Colorado, 389 UCB, Boulder, CO 80309-0389, USA}

\author[0000-0001-6947-5846]{Luis C. Ho}
\affiliation{Kavli Institute for Astronomy and Astrophysics, Peking University, Beijing 100871, China; Department of Astronomy, School of Physics, Peking University,
Beijing 100871, People's Republic of China}

\author[0000-0003-1420-6037]{Jonathan Cohn}
\affiliation{George P. and Cynthia Woods Mitchell Institute for Fundamental Physics and Astronomy, 4242 TAMU, Texas A\&M University, College Station, TX, 77843-4242,}

 \correspondingauthor{Kyle M. Kabasares}
 \email{kkabasar@uci.edu}



\begin{abstract}
We present Atacama Large Millimeter/submillimeter Array (ALMA) Cycle 2 observations of CO(2-1) emission from the circumnuclear disks in two early-type galaxies, NGC 1380 and NGC 6861. The disk in each galaxy is highly inclined ($i\,{\sim}\,75^{\circ}$), and the projected velocities of the molecular gas near the galaxy centers are ${\sim}300\,\mathrm{km \, s^{-1}}$ in NGC 1380 and ${\sim}500\,\mathrm{km \, s^{-1}}$ in NGC 6861. We fit thin disk dynamical models to the ALMA data cubes to constrain the masses of the central black holes (BHs). We created host galaxy models using Hubble Space Telescope images for the extended stellar mass distributions and incorporated a range of plausible central dust extinction values. For NGC 1380, our best-fit model yields $ \mbh = 1.47 \times 10^8\,M_{\odot}$ with a ${\sim}40\%$ uncertainty. For NGC 6861, the lack of dynamical tracers within the BH's sphere of influence due to a central hole in the gas distribution precludes a precise measurement of $\mbh$. However, our model fits require a value for $\mbh$ in the range of $(1-3) \times 10^9\,M_{\odot}$ in NGC 6861 to reproduce the observations. The BH masses are generally consistent with predictions from local BH-host galaxy scaling relations. Systematic uncertainties associated with dust extinction of the host galaxy light and choice of host galaxy mass model dominate the error budget of both measurements. Despite these limitations, the measurements demonstrate ALMA's ability to provide constraints on BH masses in cases where the BH's projected radius of influence is marginally resolved or the gas distribution has a central hole.

\end{abstract}


%\keywords{galaxies: bulges – galaxies: individual (NGC 1380, NGC 6861) – galaxies: kinematics and dynamics –
% galaxies: nuclei}

%% You can use LaTeXs \ref and \label commands to keep track of

\section{Introduction}
Supermassive black holes (BHs) are believed to reside in the centers of all massive galaxies that are not pure disks. They encode information about the formation and evolution of their host galaxies through a number of scaling relations such as the $\mbh-\sigma_{\star}$, $\mbh - L_{\mathrm{bul}}$, and $\mbh - M_{\mathrm{bul}}$ relations \citep{1995ARA&A..33..581K,2000ApJ...539L...9F,2000ApJ...539L..13G,2013ARAA..51..511K} which relate BH mass to spheroid stellar velocity dispersion, luminosity, and mass. These scaling relations are often used to estimate BH masses in galaxies over broad ranges in both galaxy type and distance \citep{2013ARAA..51..511K,2013ApJ...764..184M,2016ApJ...818...47S,2016ApJ...819...11V}.

Increasing the sample of measured BH masses that define these scaling relations is necessary to improve our understanding of BH-host galaxy coevolution. Questions regarding when these scaling relations came to be, the ranges of BH masses and galaxy types they apply to, and the physical mechanisms that allowed them to form remain unanswered. To complicate matters, $M_{\mathrm{BH}}$ predictions from the $M_{\mathrm{BH}}-\sigma_{\star}$ and $\mbh - L_{\mathrm{bul}}$ relations can strongly disagree, such as in the case of the most luminous early-type galaxies (ETGs), where the discrepancies reach an order of magnitude at $\mbh\, {\sim} 10^{10} \,M_{\odot}$ \citep{2007AJ....133.1741B,2007ApJ...662..808L}.
A larger sample of reliable BH mass measurements is needed to address these questions and issues.

About 100 supermassive BH mass measurements have been obtained through modeling the motions of either stars or gas \citep{2013ARAA..51..511K}. Accurately determining these masses requires modeling the motions of kinematic tracers that extend within the BH's radius of influence, $\rg \approx GM_{\mathrm{BH}}/\sigma_{\star}^2$, where the BH as opposed to the host galaxy is the main contributor to their combined gravitational potential. Among current measurements, the most robust are of the Milky Way's own supermassive BH \citep{2016ApJ...830...17B,2019A&A...625L..10G}, which use observations of individual stellar orbits, and of BHs in galaxies with rotating H$_2$O megamaser disks whose emission originates deep within $r_g$ \citep{1995Natur.373..127M,2011ApJ...727...20K,2018ApJ...859..172K}. 
For other galaxies, BH masses have been measured primarily through stellar-dynamical and ionized gas-dynamical modeling. Both methods are prone to a variety of challenges in modeling and interpreting the data. Stellar-dynamical modeling of massive ETGs is sensitive to the treatment of galaxy triaxiality, dark matter halo structure, and stellar orbital anisotropy \citep{2008MNRAS.385..647V,2009ApJ...700.1690G,2010ApJ...711..484S}, while ionized gas-dynamical modeling can be affected by non-circular motions and substantial gas turbulence \citep{2001ApJ...555..685B,2006MNRAS.370..559S}.

Molecular gas has emerged as a dynamical tracer capable of circumventing the aforementioned issues with stellar-dynamical and ionized gas-dynamical modeling. Tracers such as H$_2$, HCN, HCO$^{+}$, and CO emission lines have been used to constrain BH masses in late-type galaxies \citep{2007ApJ...671.1329N,2013MNRAS.429.2315S, 2015ApJ...809..101D,2015ApJ...806...39O}. In addition, a number of CO surveys have shown that a fraction of ETGs have dynamically cold and regularly rotating molecular gas disks at their centers \citep{2007MNRAS.377.1795C, 2011MNRAS.414..940Y,2013MNRAS.432.1796A,2017ApJ...846..159B}. These molecular gas disks are ideal targets for precision BH mass measurements. As with ionized gas, modeling the dynamics of molecular gas on scales comparable to the BH's sphere of influence is insensitive to factors such as the distribution of dark matter and triaxial structure, which affect stellar-dynamical models. Molecular gas also has the added benefit that it is much less turbulent than ionized gas \citep{2013MNRAS.429..534D,2015ApJ...803...16U,2017ApJ...845..170B}. 

\cite{2013Natur.494..328D} demonstrated the potential of molecular gas as an effective kinematic tracer by measuring the mass of the BH in NGC 4526 with the Combined Array for Millimeter-wave Astronomy (CARMA). Since then, the Atacama Large Millimeter/submillimeter Array (ALMA) has emerged as the premier radio interferometer for these types of measurements. There are now several ALMA-based BH mass measurements derived from observations of molecular gas emission on scales comparable to and even within the BHs' spheres of influence for nearby galaxies \citep{2016ApJ...822L..28B,2017MNRAS.468.4663O, 2017MNRAS.468.4675D,2019ApJ...881...10B,2019MNRAS.490..319N,2019MNRAS.485.4359S,2021arXiv210407779C,2021ApJ...908...19B,2021MNRAS.504.4123N}. 

In this paper, we analyze ALMA observations of NGC 1380 and NGC 6861, two galaxies that have been shown to host a central circumnuclear gas disk. \cite{2017ApJ...845..170B} mapped the distribution of CO(2-1) emission within these disks and found that they exhibited dynamically cold rotation. 

NGC 1380 is classified as an SA0 galaxy in both the Third Reference Catalogue of Bright Galaxies (RC3; \citealp{1991rc3..book.....D}) and in the Hyperleda database \citep{2003A&A...412...45P}. It is located at a luminosity distance of 17.1 Mpc in the Fornax cluster based on surface-brightness fluctuations from \citet{2001ApJ...546..681T} after applying the Cepheid zero-point correction from \citet{2007ApJ...655..144M}. With this assumed luminosity distance, and using an observed redshift of $z = 0.00618$ obtained from initial dynamical modeling results, the corresponding angular scale is 81.9 $\mathrm{pc}\,\mathrm{arcsec}^{-1}$. We adopt the Hyperleda average stellar velocity dispersion of $\sigma_{\star} = 215\,\mathrm{km}\,\mathrm{s^{-1}}$ \citep{2014A&A...570A..13M}, a total apparent $K$-band magnitude of $m_K = 6.87$ mag from the Two Micron All Sky Survey (2MASS; \citealp{2003AJ....125..525J})  and a bulge-to-total ratio of $B/T = 0.34$ from the Carnegie-Irvine Galaxy Survey (CGS; \citealp{2019ApJS..244...34G}).

NGC 6861 is located at a luminosity distance of 27.3 Mpc in the Telescopium galaxy group, which  corresponds to an angular scale of 129.9 $\mathrm{pc}\,\mathrm{arcsec}^{-1}$ when using a redshift of $z = 0.00944$ from our initial dynamical models. This galaxy is classified as an E/S0 in Hyperleda and as an S0A-(s) in RC3. \cite{2013ARAA..51..511K} note that the main body of NGC 6861 does not deviate significantly from an $n \simeq 2$ Sérsic-function profile, although the galaxy has extra central light; we adopt that paper's classification as an extra-light elliptical. In addition, we also adopt the stellar velocity dispersion of $\sigma_{\star} = 389\,\mathrm{km}\,\mathrm{s^{-1}}$ measured within the effective radius by \cite{2013AJ....146...45R}, and the 2MASS total apparent $K$-band magnitude of $m_K = 7.75$ mag \citep{2016ApJ...818..182V}.

For each galaxy, we obtained a BH mass by constructing gas-dynamical models and fitting directly to the ALMA data cubes. This work provides the first dynamical mass measurement of the supermassive BH in NGC 1380, and a second, independent dynamical mass measurement of the supermassive BH in NGC 6861, which was previously measured through stellar-dynamical modeling to be $(2.0 \pm 0.2) \times 10^9\,M_{\odot}$ by \cite{2013AJ....146...45R}. 

Our paper is organized as follows. In Section \ref{sec:Observations}, we describe both the Hubble Space Telescope (HST) and ALMA observations and the data reduction process. In Section \ref{sec:HostGalaxySurfaceBrightness}, we decompose the stellar surface brightness distribution of each galaxy with Multi-Gaussian Expansions and construct extinction-corrected host galaxy models. A description of our dynamical modeling formalism, including the thin disk model, is presented in Section \ref{sec:Dynamical Modeling}. We present and compare the results of our dynamical models in Section \ref{sec:Results}. In Section \ref{sec:Discussion}, we compare our measurements of $\mbh$ to the local BH-host galaxy scaling relations and discuss limiting factors in the measurements. We conclude and summarize our findings in Section \ref{sec:Conclusion}.

\section{Observations}
\label{sec:Observations}
\subsection{HST Data}
\label{sec:HSTObservations}
For NGC 1380, we retrieved and used archival HST F160W ($H$-band) images from HST program 11712. The observation was subdivided into 4 separate exposures of 299 seconds each that were taken with the Wide Field Camera 3 (WFC3). We  processed the images with the \texttt{calwf3} pipeline and subsequently combined them in \texttt{AstroDrizzle} to produce a cleaned and distortion-corrected image with a pixel scale of $0\farcs08$ $\mathrm{pixel}^{-1}$. We followed a similar procedure for NGC 6861. We used archival HST data for NGC 6861 from Program 15226, which was designed to obtain host galaxy imaging to complement our ALMA program. The observation consisted of 4 separate exposures of 249 seconds each taken with the F160W filter on WFC3. We processed and combined the images with \texttt{calwf3} and \texttt{AstroDrizzle} to produce a composite image with a $0\farcs08$ $\mathrm{pixel}^{-1}$ scale. To identify regions of substantial dust attenuation, we also obtained archival F110W ($J$-band) images for each galaxy. Archival F110W observations of NGC 1380 were obtained from HST program 11712 and consisted of 4 separate exposures of 299 seconds each, while F110W observations of NGC 6861 were obtained from HST program 15226 and consisted of 2 separate exposures of 249 seconds each. These were processed using the \texttt{calwf3} pipeline and \texttt{AstroDrizzle} as above.

\subsection{ALMA Data} 
\label{sec:ALMAObservations}
\subsubsection{Observations and Data Reduction}
We obtained ALMA imaging of NGC 1380 and NGC 6861 as part of Program 2013.1.00229.S. The data were studied by \cite{2017ApJ...845..170B} as part of a larger sample of galaxies to map CO(2-1) emission in nearby ETGs, and we review their data reduction process and findings below.

NGC 1380 was observed in ALMA Band 6 for 23 minutes on both 2015 June 11 and 2015 September 18 with maximum baselines of 783 m and 2125 m, respectively. For the redshifted $^{12}\mathrm{CO}(2-1)$ line, the observation covered a 1.875 GHz bandwidth from 228.199 GHz to 230.074 GHz, centered at an estimated redshifted line frequency of 229.136 GHz. The frequency channel widths were 488.281 kHz, corresponding to a velocity channel resolution of 0.64 km $\mathrm{s^{-1}}$ at the redshifted frequency. For continuum emission, two separate 2 GHz spectral windows were centered at 227.210 GHz and 244.902 GHz with 15.625 MHz channel widths, equating to velocity resolutions of 20.6 km $\mathrm{s^{-1}}$ and 19.1 km $\mathrm{s^{-1}}$, respectively. The data were initially processed through the ALMA pipeline with version 4.3.1 of the Common Astronomy Software Applications package (CASA,  \citealp{2007ASPC..376..127M}) and then imaged into data cubes using Briggs weighting with a robust parameter of 0.5 following continuum phase self-calibration and continuum subtraction in the $uv$ plane. We reimaged the cube to have 10 km $\mathrm{s^{-1}}$ velocity channel widths (with respect to the rest frequency of the $^{12}\mathrm{CO}(2-1)$ line) to isolate narrrower line features in spatial regions close to the disk center and chose a pixel size of $0\farcs{03}$ to sufficiently sample the synthesized beam's minor axis. The beam's full width at half maximum (FWHM) is $0\farcs{24}$ and $0\farcs{18}$ along the major and minor axis, respectively, and the beam has a position angle of $86.9^{\circ}$ measured east of north.

NGC 6861 was observed on 2014 September 01 in ALMA Band 6 for 24 minutes with a maximum baseline of 1091 m. The line observation was centered at an estimated redshifted $^{12}\mathrm{CO}(2-1)$ line frequency of 228.390 GHz, while the continuum windows were centered at 226.466 GHz and 244.098 GHz, with the same bandwidth and channel spacing properties as the NGC 1380 observation. The data were processed using CASA version 4.2.2 and imaged into a data cube with 20 km $\mathrm{s^{-1}}$ velocity channel widths following the continuum phase self-calibration and continuum subtraction processes described for the NGC 1380 data. The NGC 6861 data cube has the following properties: a synthesized beam size of $0\farcs{32} \times 0\farcs{23}$ with a position angle of 58.2$^{\circ}$ and a pixel size of $0\farcs{065}$.

\begin{figure}
    \centering
    \includegraphics[width=3.1in]{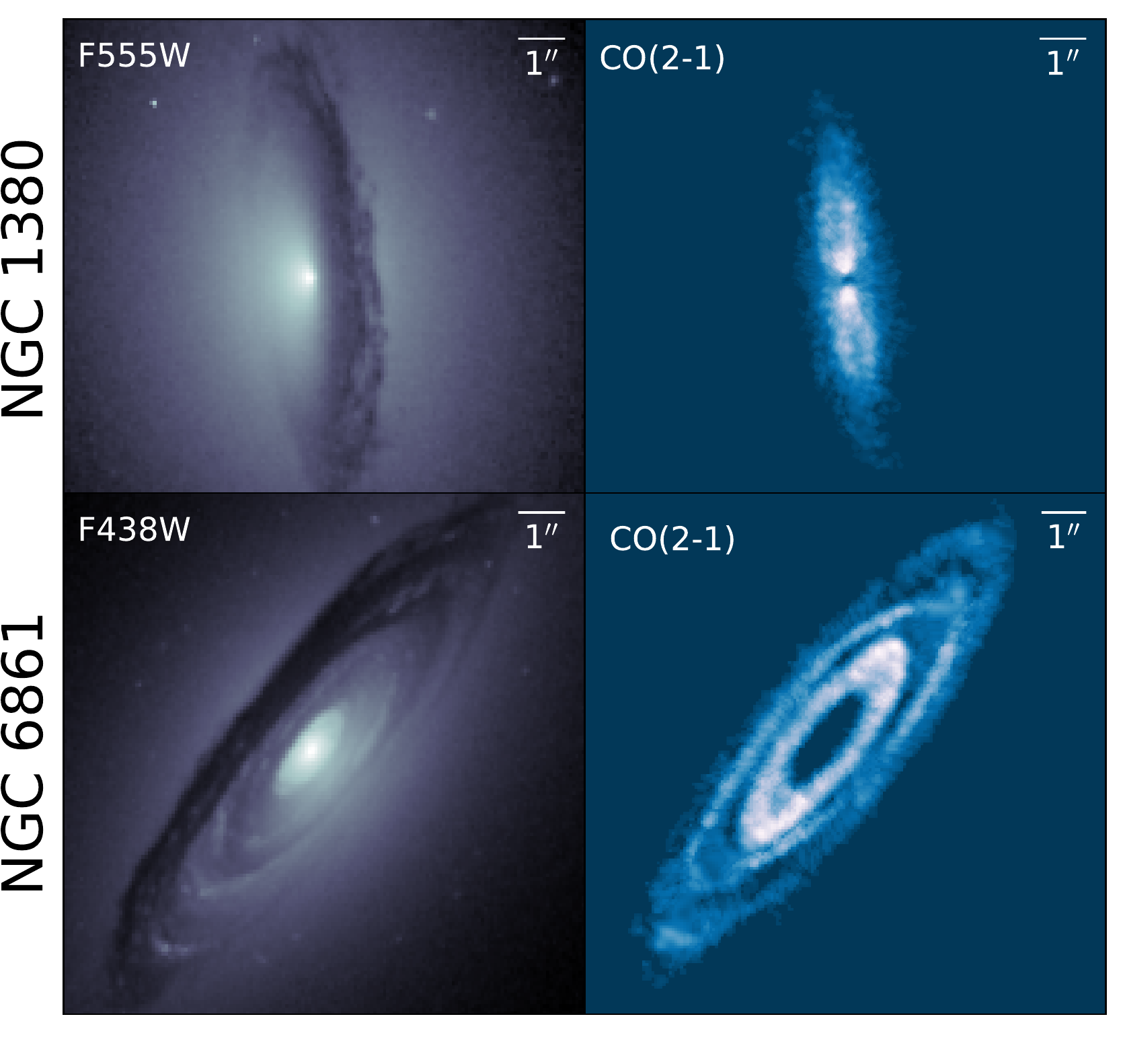}
    \caption{Images of NGC 1380 and NGC 6861 from HST and ALMA observations showing the co-spatial distributions of the dust and gas. The left panels show F555W and F438W HST observations of the dust disks in NGC 1380 and NGC 6861, respectively. For each image, North is up and East is to the left. ALMA intensity maps in the right-side  panels were created by summing across channels after using the 3DBarolo program \citep{2015MNRAS.451.3021D} to generate a mask that identified pixels with CO emission.} 
    \label{fig:hst_alma_observations}
\end{figure}

\begin{figure*}[ht]
    \centering
    \includegraphics[width=3in]{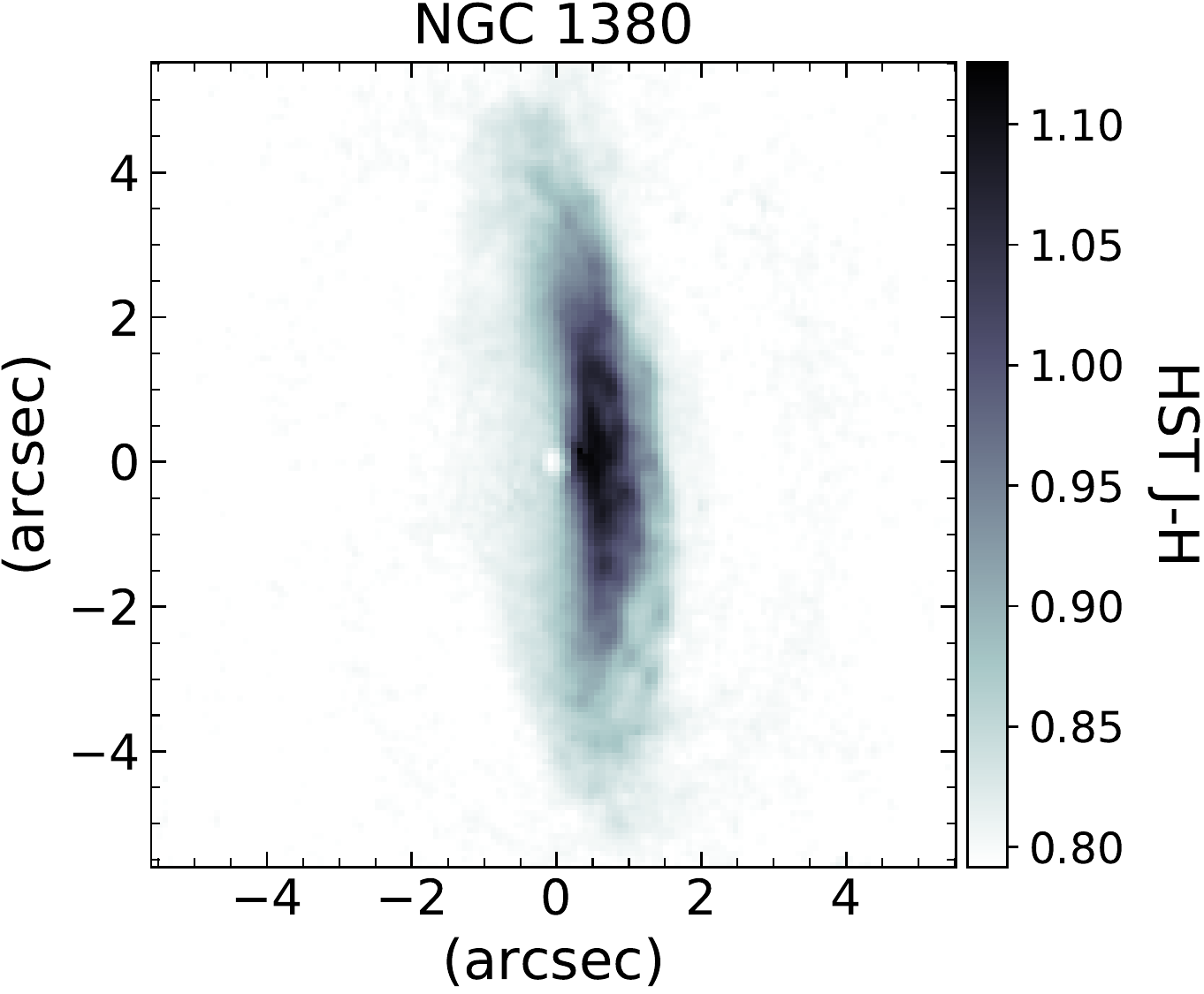}
    \hspace{5mm}
    \includegraphics[width=3in]{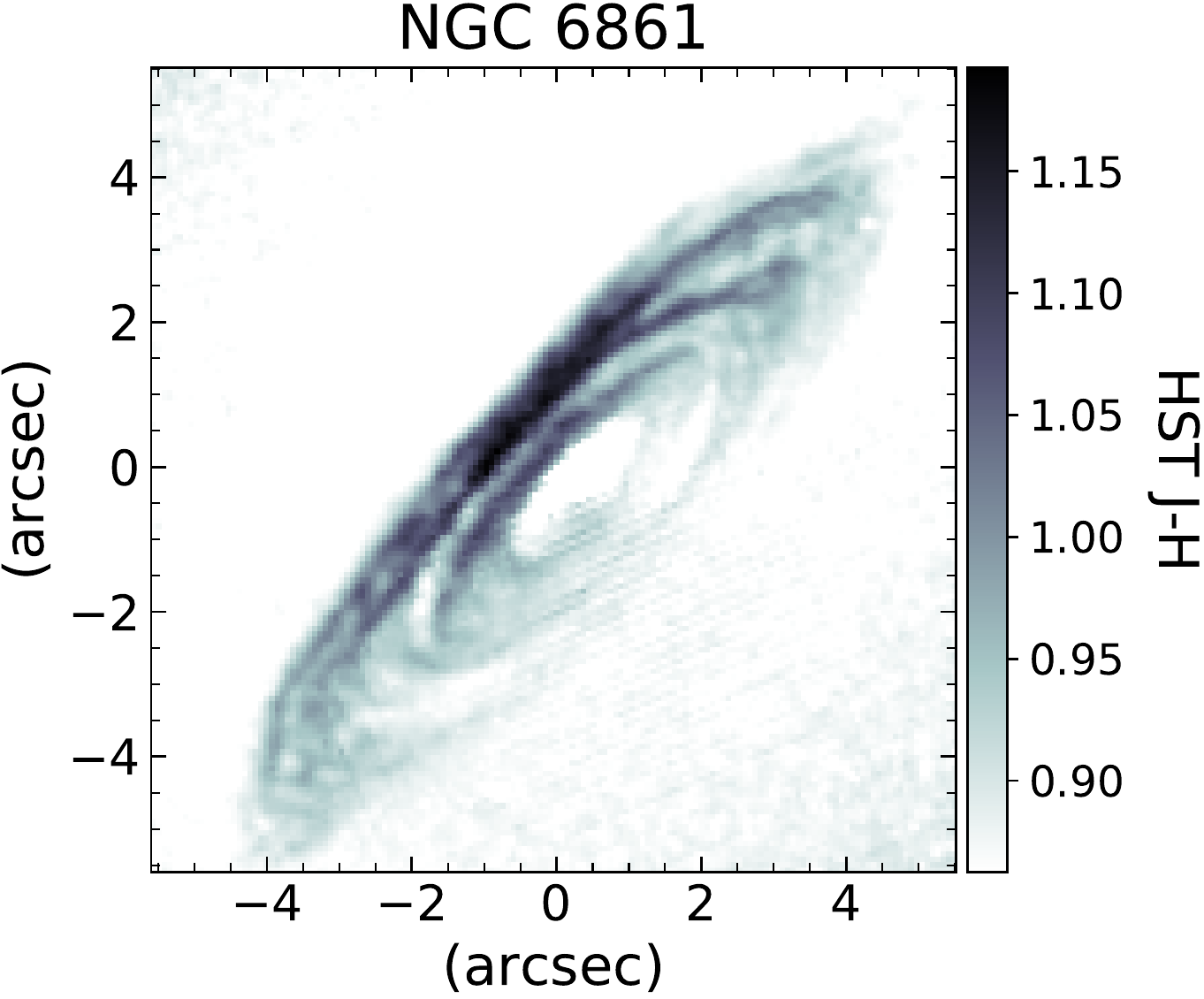}
    \caption{NGC 1380 (left) and NGC 6861 (right) $J-H$ color maps constructed using WFC3 F110W and F160W observations. The maps highlight the color asymmetries of the near and far sides of the disks, with the near sides being ${\sim}0.2-0.3$ mag redder than the far sides. For NGC 6861, a 1$\arcsec$ radius hole in the center of the dust distribution can be seen.}
    \label{fig:jminushcolor}
\end{figure*}

\subsubsection{Circumnuclear Disk Properties}
\cite{2017ApJ...845..170B} determined several properties of the circumnuclear disks in NGC 1380 and NGC 6861, which we summarize here. The gas in each disk is co-spatial with the dust, as seen in the HST optical images and ALMA integrated intensity maps in Figure \ref{fig:hst_alma_observations}. Both disks are very inclined ($i \approx 75^{\circ}$) and exhibit orderly rotation around their respective centers, with projected line-of-sight (LOS) velocities of $\sim$300 km $\mathrm{s}^{-1}$ and $\sim$500 km $\mathrm{s}^{-1}$ for NGC 1380 and NGC 6861, respectively. LOS velocity and dispersion maps indicate nearly circular and dynamically cold rotation about the disk centers. Independent stellar kinematic observations with the Multi Unit Spectroscopic Explorer (MUSE) have also revealed the presence of a large-scale cold disk component in NGC\,1380 \citep{2018A&A...616A.121S}. The radial extents of the CO emission were measured to be 5\farcs{2} (426 pc) and 6\arcsec{} (784 pc) for NGC 1380 and NGC 6861, respectively. A major axis position-velocity diagram (PVD) extracted from the NGC 1380 data cube shows a slight rise in velocity within the innermost ${\sim}0\farcs{1}$, although this does not extend past the velocities observed in the outer parts of the PVD. This central upturn in gas velocity indicates the presence of a massive and compact object at the disk center. For NGC 6861, the PVD and moment maps reveal a central ${\sim}1\arcsec{}$ radius hole in CO emission. The gas mass of each disk was determined by summing the CO flux and assuming an $\alpha_{\mathrm{CO}}$ factor of $3.1\, M_{\odot} \,\, \mathrm{pc}^{-2}$ ($\mathrm{K}\, \mathrm{km}\,\mathrm{s}^{-1})^{-1}$ \citep{2013ApJ...777....5S} as the extragalactic mass-to-luminosity ratio, a CO(2-1)/CO(1-0) $\approx 0.7$ line ratio in brightness temperature units \citep{1999AJ....117.1995L}, and a correction factor of 1.36 for helium. Given these assumptions, the gas masses were estimated to be $(8.4 \pm 1.6) \times 10^7 \,M_{\odot}$ and $(25.6 \pm 8.9) \times 10^7 \,M_{\odot}$ for NGC 1380 and NGC 6861, respectively. For NGC\,1380, our assumption about CO excitation can be tested: \cite{2019MNRAS.483.2251Z} measure a CO(1--0) line flux for NGC\,1380 that in combination with the CO(2--1) line flux from \cite{2017ApJ...845..170B} implies a CO(2--1)/CO(1--0) ratio of $1.08^{+0.24}_{-0.20}$ in brightness temperature units. This is higher than our assumed value of 0.7, and implies a lower gas mass. Because a ratio $> 1$ is unphysically high if both CO lines are tracing the same material, we consider a value $\approx 0.9$ (still lying within the measurement uncertainties) to be more appropriate and in Section 5.2 explore the implications of the correspondingly lower gas mass for our dynamical models.

\section{Host Galaxy Surface Brightness Modeling}

\label{sec:HostGalaxySurfaceBrightness}
A key input to the dynamical modeling program is the stellar mass profile, $M_{\star}(r)$, which is determined by measuring and deprojecting the host galaxy's observed surface brightness profile. One approach to obtaining $M_{\star}(r)$ is the Multi-Gaussian Expansion (MGE) method, which fits the observed brightness in galaxy images with a series expansion of two-dimensional Gaussian functions \citep{1994A&A...285..723E,2002MNRAS.333..400C}. For galaxies such as NGC 1380 and NGC 6861, which possess optically thick dust disks, the impact of dust attenuation on the host galaxy light is mitigated by using observations in the near-infrared (NIR) regime, but the impact is not completely negligible. To assess the variation of $H$-band extinction across the dust disk, we performed a simple correction to the observed $H$-band major axis surface brightness profile for each galaxy as described in Section \ref{sec:DustCorrection}. This was done as an attempt to quantify the possible impact of dust on the host galaxy's surface brightness. We then proceeded to fit dust-masked and dust-corrected MGE models to the drizzled $H$-band images of NGC 1380 and NGC 6861, as described in Sections \ref{sec:NGC1380HostGalaxyModels} and \ref{sec:NGC6861HostGalaxyModels}.

\begin{figure*}[ht]
    \centering
    \hspace{-10mm}
    \includegraphics[width=3in]{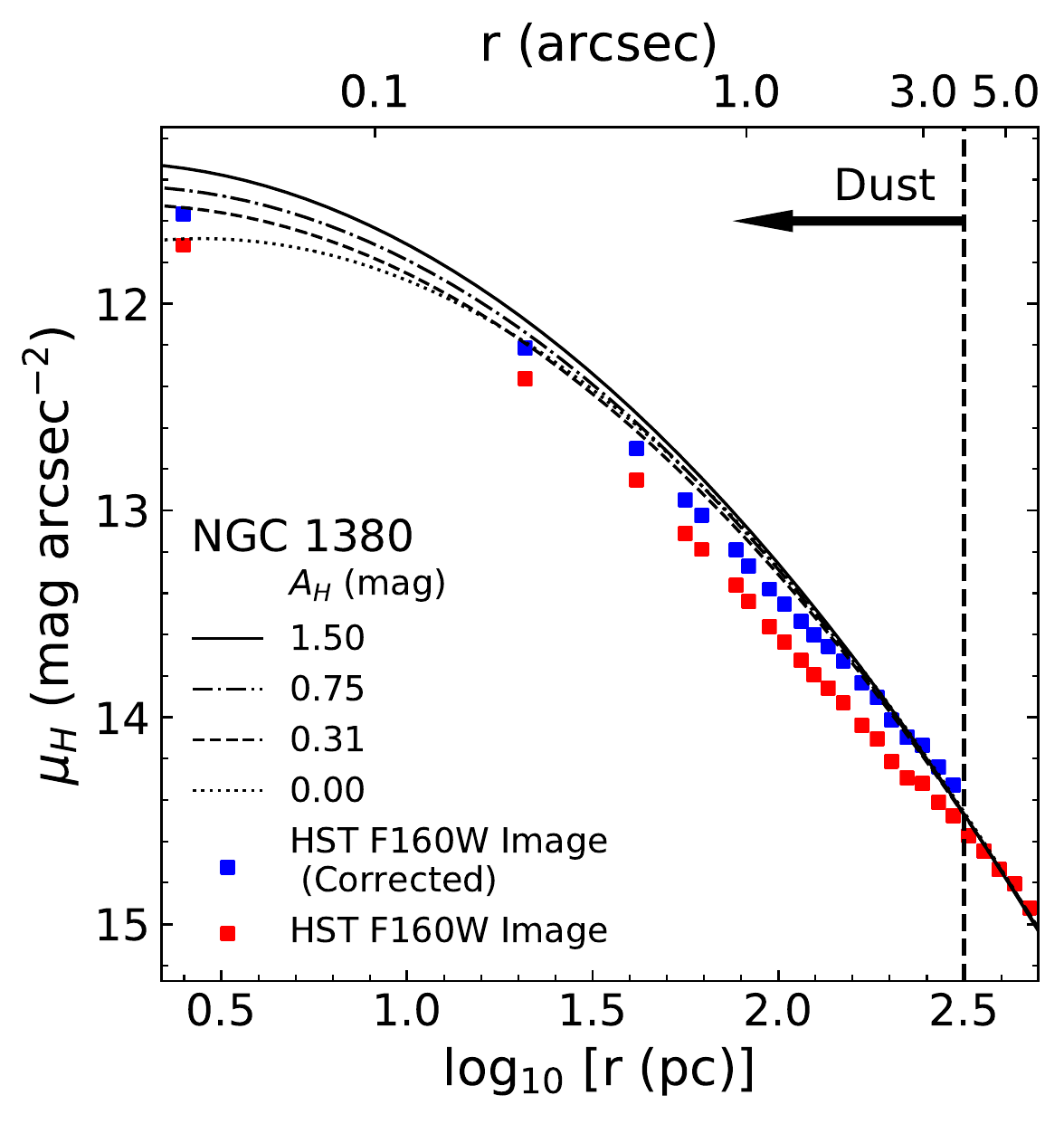}
     \hspace{5mm}
    \includegraphics[width=3in]{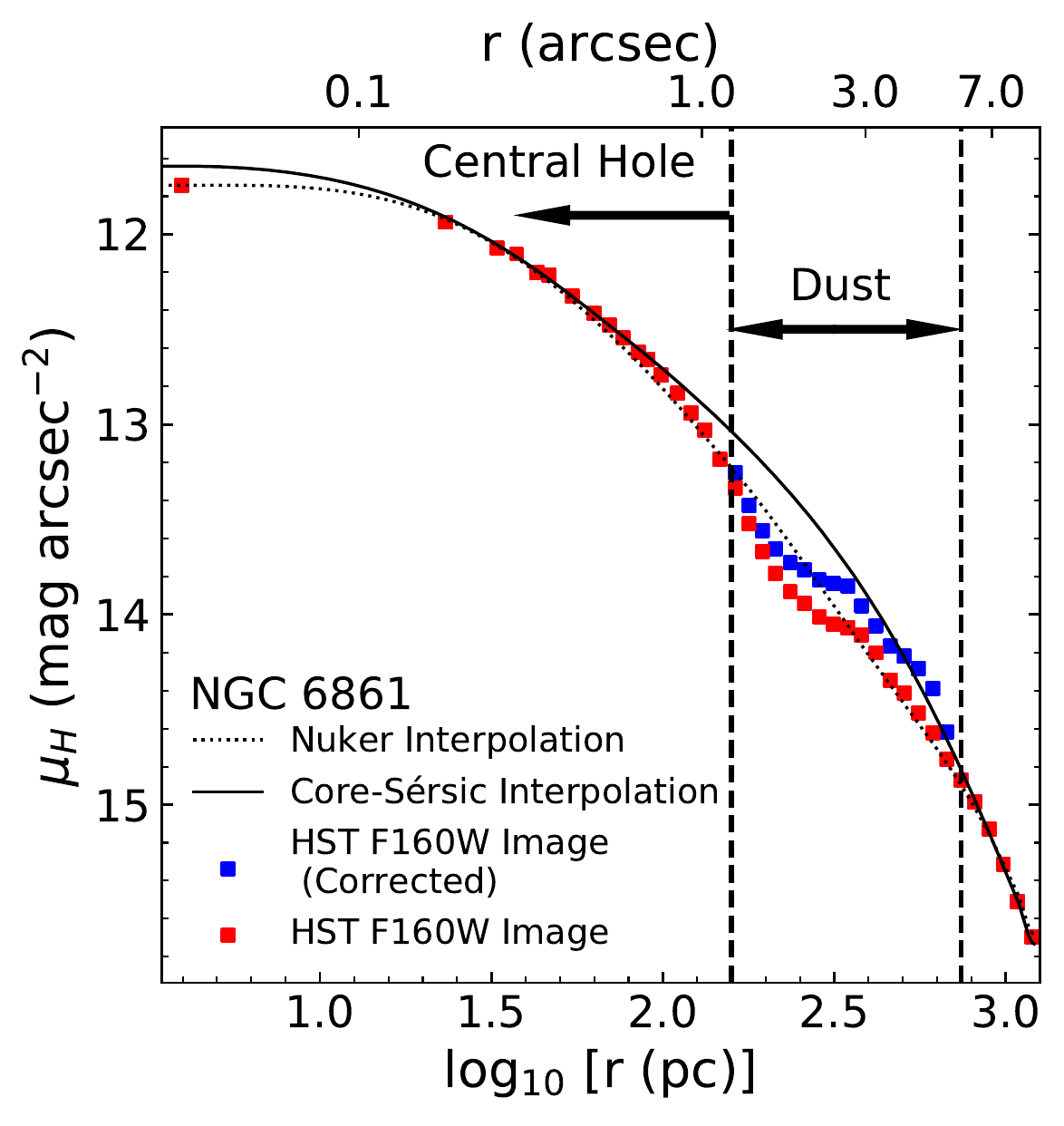}
    \caption{Comparison of the observed NGC 1380 (left) and NGC 6861 (right) HST major axis $H$-band surface brightness profiles to their respective dust-corrected models. The red points are the observed values from the $H$-band images, while the blue points are the dust-corrected values described in Section \ref{sec:DustCorrection}. The different lines in each panel correspond to extracted major axis surface brightness profiles for our 2D MGE models, which are described in Sections \ref{sec:NGC1380HostGalaxyModels} and \ref{sec:NGC6861HostGalaxyModels}. For NGC 1380, we mark the outer edge of the dust disk with a vertical dashed line and indicate that the dust extends down to the nucleus with an arrow, while for NGC 6861, we indicate the inner and outer boundaries of the dusty region with the two dashed lines.}
    \label{fig:dustcorrections}
\end{figure*}

\subsection{Major Axis Dust Extinction Corrections}
\label{sec:DustCorrection}
The central dust disk in each galaxy is clearly visible in Figures \ref{fig:hst_alma_observations} and \ref{fig:jminushcolor}, due to its dimming and reddening of the observed stellar light. We attempted to estimate the amount of dust extinction with a color-based correction method, which was complicated by the fact that the dust disk is embedded within the galaxy and cannot be treated as a simple foreground screen. To estimate the amount of dust extinction, we extracted and corrected the observed $H$-band major axis surface brightness profile of each galaxy. Using the \texttt{sectors\_photometry} routine from the \texttt{MgeFit} package in Python \citep{2002MNRAS.333..400C}, we plotted the surface brightness profile of each galaxy in Figure  \ref{fig:dustcorrections}. In NGC 1380, a slight dip in the profile can be seen at around $r = 4\arcsec{}$, which marks the outer edge of the dust disk, while a more noticeable dip is seen between ${\sim}1\arcsec{}-5\farcs{5}$ in NGC 6861.

To determine the pixels that are most affected by dust, we created $J-H$ color maps as seen in Figure \ref{fig:jminushcolor}. We note that all $H$ and $J$-band magnitudes in this work are in the Vega magnitude system. The color maps revealed that the dust extends from the nuclei and the pixels most affected by dust are about 0.25 mag redder than the median $J-H$ color of ${\sim}$0.80 mag outside the disk. Furthermore, each nucleus has a bluer color that is nearly identical to the median color outside the disk. A blue nucleus suggests the presence of star formation or a weak active galactic nucleus (AGN); we discuss these possibilities below. 

\begin{figure}
    \includegraphics[width=3.2in]{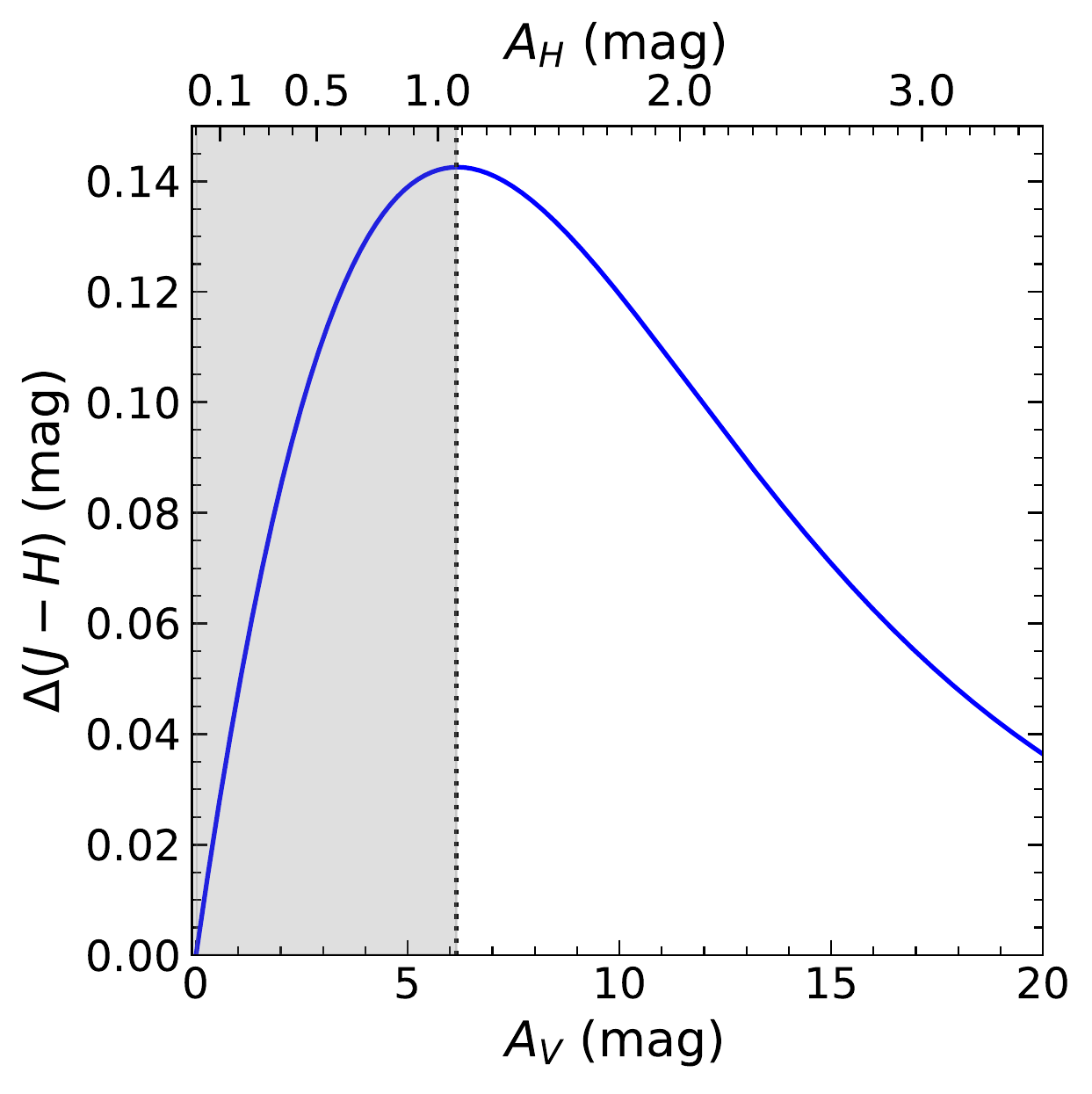}
    \caption{Modeled $\Delta(J-H)$ curve as a function of $A_V$ (bottom axis) and $A_H$ (top axis). The curve was generated using equation 1 from \cite{2019ApJ...881...10B} and a standard Galactic ($R_V = 3.1$) extinction curve \citep{1985ApJ...288..618R}, under the assumptions that along the major axis, the fractions of light in front of and behind the disk are equal $(f = b = 0.50)$, and that only the starlight originating from behind the disk is subject to dust extinction. The gray region left of the dotted line indicates the part of the curve we used to define a one-to-one correspondence between $\Delta(J-H)$ and $A_V$ and $A_H$ for the low-extinction branch of the curve. }
    \label{fig:colorexcesscurve}
\end{figure}

We attempted to correct the major axis $H$-band surface brightness profiles by examining $\Delta(J-H)$, the observed color excess relative to the median $J-H$ color outside the disk, along the major axis. Using equation 1 from \cite{2019ApJ...881...10B}, which predicts the ratio of observed to intrinsic integrated stellar light based on the embedded-screen model described by  \cite{2017MNRAS.472.1286V}, we generated a model $\Delta(J-H)$ curve as a function of intrinsic $V$-band extinction, $A_V$, to compare with the observations. 

The embedded-screen model assumes that the obscuring dust lies in a thin, inclined disk that bisects the galaxy, and that the fraction of stellar light originating behind the disk, $b$, is obscured by simple screen extinction, while the fraction of stellar light in front of the disk, $f$, is unaffected. In addition, the model assumes that there is no scattering of stellar light back into the LOS, and that the $J-H$ color outside of the disk is the intrinsic color of the host galaxy. One important aspect of this color excess model is that $\Delta(J-H)$ is not a strictly monotonically increasing function of intrinsic extinction. As seen in Figure \ref{fig:colorexcesscurve}, the color excess increases approximately linearly with increasing $A_V$ up to a turnover point, after which it will begin to decrease to zero as the light originating behind the disk becomes completely obscured. As a result, there are two possible $A_V$ values for a given $\Delta(J-H)$. To maintain a one-to-one correspondence between $\Delta(J-H)$ and $A_V$, we considered only the low extinction branch of the curve and therefore adopted the lesser of the two possible $A_V$ values. We inverted the relationship to derive $A_V$ as a function of observed $\Delta(J-H)$ by fitting a third-order polynomial up to the turnover point and determining its inverse. Using a standard Galactic ($R_V = 3.1$) extinction curve \citep{1985ApJ...288..618R} where $A_H/A_V = 0.175$, and assuming that the fractions of stellar light originating in front of and behind the disk are equal ($f = b = 0.5)$ along the major axis, we associated our observed major axis $\Delta(J-H)$ values with corresponding values of $A_H$. 

We generated point by point corrections to our major axis surface brightness profiles using the fact that our modeled $A_H$ values 
only applied to the fraction of light originating behind the disk. The corrected values are shown in blue in Figure  \ref{fig:dustcorrections} for points within the dust disk for each galaxy.  For NGC 1380, the corrected surface brightness profile still exhibits a slight dip near the edge of the dust disk. For NGC 6861, where we can anchor the surface brightness profile inside and outside the dust disk, it is clear there is still some remaining extinction, as the decrease in observed surface brightness is still visible in the corrected profile. 

Since the method described above appears to undercorrect for extinction, we also tried applying the method using the high-extinction branch of the $\Delta(J-H)$ vs. $A_V$ curve to both galaxies; however, this approach led to overcorrection all  along the major axis of each galaxy, as each point's $H$-band surface brightness was raised by nearly the theoretical maximum of 0.75 mag arcsec$^{-2}$. While our method provides some insight on how extinction varies across the disk, based on the results for these two galaxies using both branches of the color excess curve, it is clear that a simple extinction correction for a thin embedded disk does not fully correct for dust extinction or give us accurate host galaxy profiles. Thus, we opted to create dust-masked and dust-corrected MGEs to model the host galaxy's light following methods used for similar galaxies by \cite{2019ApJ...881...10B,2021ApJ...908...19B} and \cite{2021arXiv210407779C}.

\begin{figure}[b]
    \centering
    \includegraphics[width=3in]{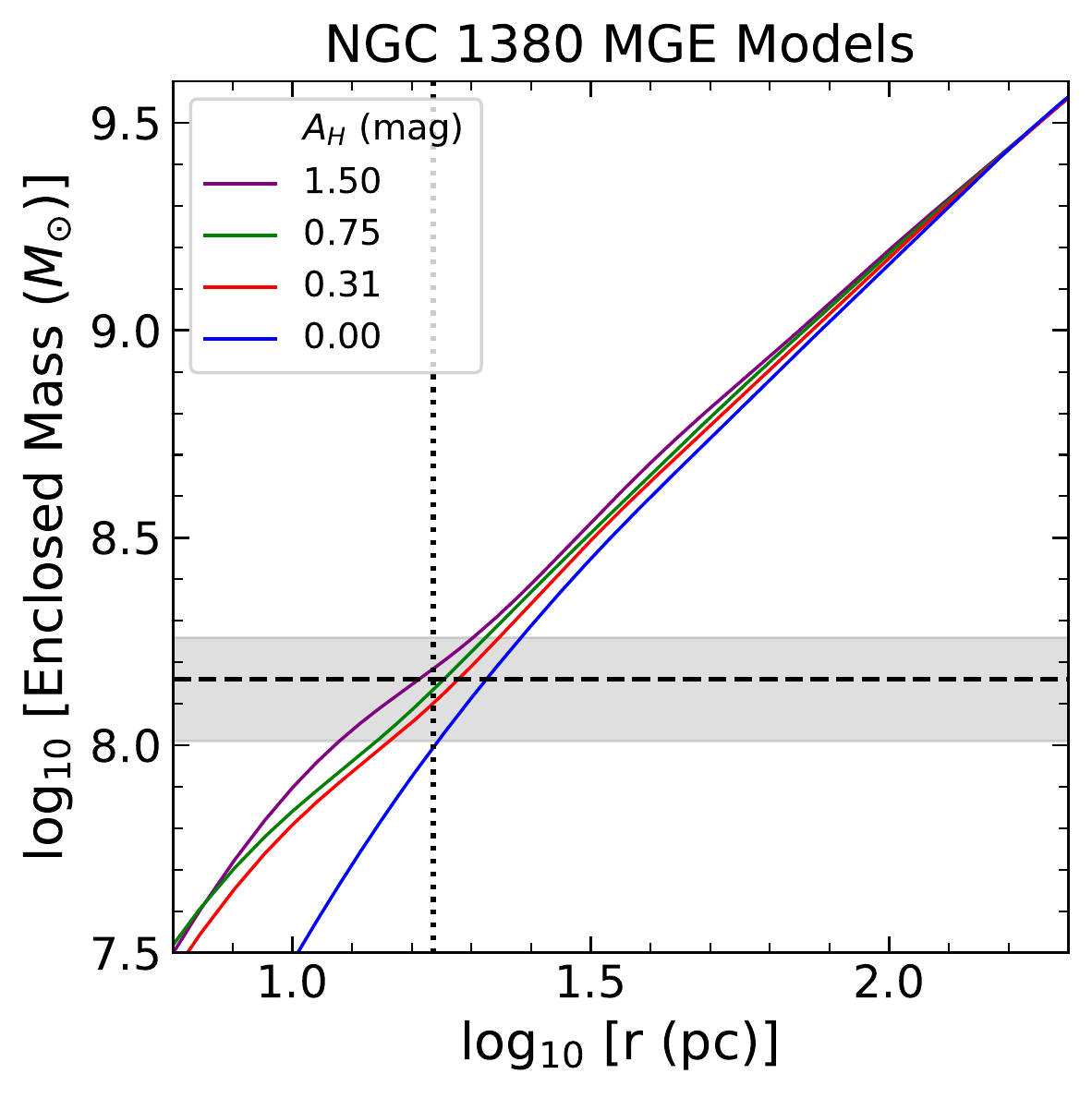}
    \caption{Plot of $\log_{10} M_{\star}(r) $ vs. $\log_{10} r$ in NGC 1380 for the four different MGE models, determined by calculating $M_{\star}(r) = rv_{\star,\,\mathrm{MGE}}^2/G$, where the $v_{\star,\,\mathrm{MGE}}$ values have been scaled by their respective $\sqrt{\Upsilon_H}$ values in Table \ref{tabledynparams}. The resolution of the ALMA observation is denoted by the vertical dotted line and is comparable to the BH's expected radius of influence. The BH mass determined from our fiducial model is represented by the horizontal dashed line, and the range of BH masses determined from Models A-D is indicated by the gray shaded region.}
    \label{fig:ngc1380MGEenclosedmass}
\end{figure}

\begin{figure*}[t]
    \centering
    \includegraphics[scale=0.70]{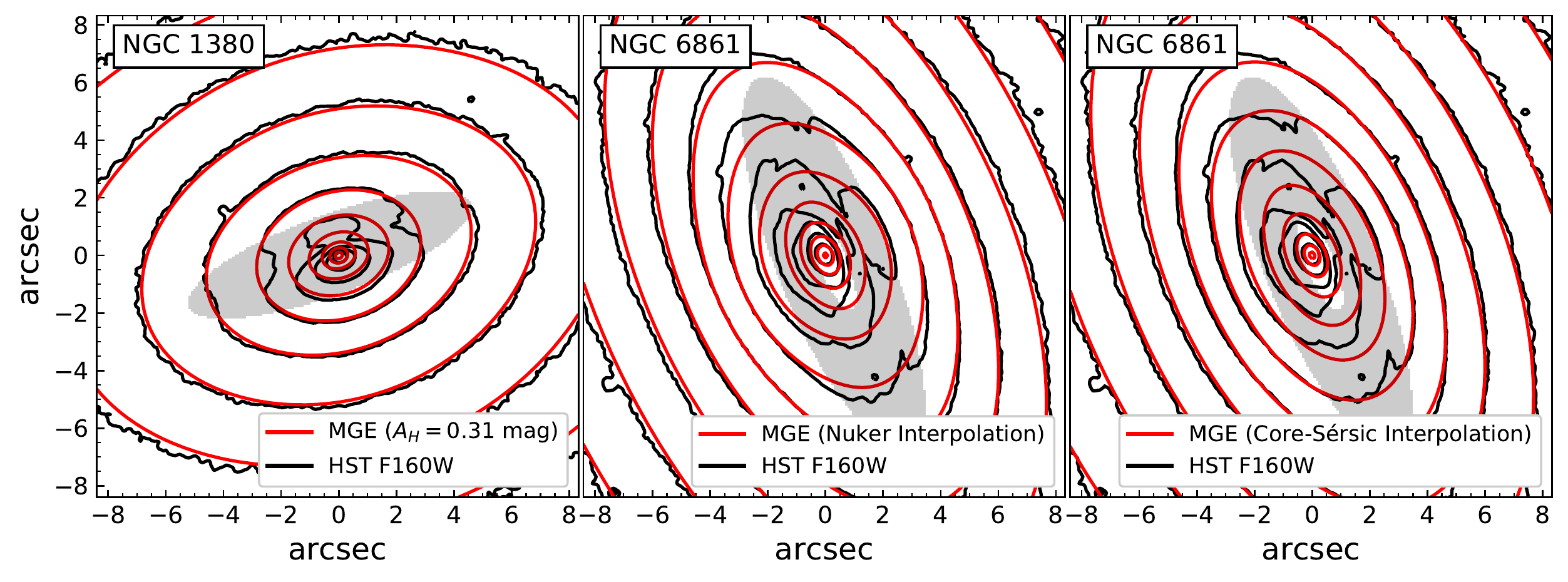}
    \caption{Isophotal contour maps of the HST F160W images of NGC 1380 and NGC 6861 displayed on an $8\arcsec{}\times8\arcsec{}$ scale. For NGC 1380 (left panel), we superpose the surface brightness contours of the intrinsic $A_H = 0.31$ mag MGE model, while for NGC 6861 (center and right panels), we display the contours of both the Nuker and Core-Sérsic interpolation MGE models. The shapes and sizes of the central dust disks are indicated by the shaded gray ellipses.}
    \label{fig:mgecontours}
\end{figure*}

\subsection{NGC 1380 Host Galaxy Models}
\label{sec:NGC1380HostGalaxyModels}
Before we constructed an MGE model for NGC 1380, we created a mask that isolated the host galaxy light from the light of foreground stars and background galaxies and identified pixels affected by dust. For the NGC 1380 drizzled $H$-band image, we masked out foreground stars and background galaxies in the image and corrected for a foreground Galactic reddening in the $H$-band of $A_H = 0.009$ mag based on reddening measurements from Sloan Digital Sky Survey data by \cite{2011ApJ...737..103S}. Using the $J-H$ color map we constructed earlier, we also masked pixels that had $J-H$ $> 1.05$ mag, which were on the disk's near side. This step prevented pixels with the most apparent dust obscuration from being used in the MGE fit, but there is still clear evidence of extinction in other regions of the disk. 

We modeled the observed $H$-band surface brightness within the inner 10$\arcsec{}\times\,$10$\arcsec{}$ with an MGE created in \texttt{GALFIT} \citep{2002AJ....124..266P} and required each Gaussian component to have the same center and position angle. To account for the HST PSF, we generated a model $H$-band PSF using Tiny Tim \citep{krist04}. This PSF was drizzled and dithered in the same manner as the $H$-band image and, along with our mask, was used during the \texttt{GALFIT} optimization to create the MGE. We refer to this initial, dust-masked MGE as our $A_H = 0.00$ mag model, since it does not attempt to correct for the impact of extinction at locations that were not masked out.  

 A robust pixel-by-pixel dust correction model would require radiative transfer modeling to account for factors such as disk geometry, thickness, scattering from dust, and extinction within the disk itself \citep{2013A&A...550A..74D,2015A&C.....9...20C}. Additionally, light originating from recent star formation or a weakly active nucleus would add further complications to a dust correction model. In our $J-H$ color map, the nucleus of NGC 1380 is bluer than the most reddened pixels in our mask by about ${\sim}0.2$ mag, suggesting the presence of star formation and/or a weak AGN. \cite{2020MNRAS.496.2155Z} used combined MUSE and ALMA data to study the relationship between molecular gas surface density and star formation rate in NGC 1380. They concluded that there was no H$\alpha$ emission from star formation, and that the presence of H$\alpha$ in NGC 1380 was primarily due to what they defined as composite regions such as shocks or an AGN. Indeed, through integral field spectroscopy, \cite{2014MNRAS.440.2419R} determined that NGC\,1380 contains a low ionization nuclear emission-line region (LINER). 

\cite{2019A&A...622A..89V} also study the dust mix and gas properties in NGC 1380 with MUSE observations and detect low-level star formation within the inner portion of the disk. They construct 2D $A_V$ maps of the dust lane area by comparing MGE model fits (after having masked out the dust lane) to MUSE $V$-band images and estimated a maximum $A_V$ value of 1.00 mag, corresponding to $A_H \approx 0.18$ mag for a standard Galactic ($R_V = 3.1$) extinction curve. In addition, they use 3D radiative transfer models to reproduce the observed $V$-band attenuation (defined as the combination of extinction and scattering of light back into the LOS) curve. However, their methods assume that only the near side of the dust disk experiences any $V$-band extinction, whereas our $J - H$ map shows that pixels on the far side are redder relative to the median color outside the disk, indicating that light from the far side is still affected by extinction.

We used the simpler method described by \cite{2019ApJ...881...10B}, which assumed an analytic surface brightness profile model to correct for dust extinction. Their method examined the impact of extinction on the inferred host galaxy circular velocity profile by adjusting the central $H$-band surface brightness profile to correct for three fiducial values of dust extinction. We chose the same values of $H$-band extinction as \cite{2019ApJ...881...10B}, which were $A_H = 0.31, 0.75$, and $1.50$ mag. These values correspond to fractions of 1/4, 1/2, and 3/4 of the stellar light originating behind the dust disk. We emphasize that these values were chosen to explore the impact of dust on the inferred host galaxy models (and subsequently, the values of $\mbh$ derived from our dynamical models) over a range in extinction and note that among the major axis surface brightness profiles shown in Figure \ref{fig:dustcorrections}, the $A_H = 0.31$ mag MGE model most closely resembles the observed profile after correction using the low-extinction branch of the reddening curve.

We created three dust-corrected MGE models based on the three fiducial $H$-band extinction values mentioned above. We followed the steps outlined by \cite{2021ApJ...908...19B} and describe our process below.  To start, we fit a 2D Nuker model \citep{1997AJ....114.1771F} in \texttt{GALFIT} to the central $10\arcsec{} \times 10 \arcsec{}$ region of the $H$-band image, using the same dust mask and PSF model as before. Nuker profiles are known to effectively model the central surface brightness profiles of ETGs, and we can easily adjust their parameters to produce dust-corrected models matching the $H$-band image. The Nuker model's surface brightness profile is characterized by inner and outer power-law profiles, with $\gamma$ and $\beta$ representing the inner and outer profile logarithmic slopes. The transition between these two regimes occurs at a break radius, $r_\mathrm{b}$, and the transition sharpness is controlled by the parameter $\alpha$. We allowed all free parameters of the Nuker model to vary in this initial fit. The Nuker model parameters  converged to $\alpha = 0.42,\, \beta = 1.49, \,\gamma = 0.31$, and $r_\mathrm{b} = 2\farcs{5}$ (${\approx} 200$ pc). These parameters characterized the Nuker model fit to the $H$-band image prior to any dust correction. We then manually corrected the central surface brightness values of the $H$-band image for extinction levels of $A_H = 0.31,\,0.75$, and $1.50$ mag, and fit three separate Nuker models to these three dust-corrected images, keeping all parameters other than $\gamma$ fixed. This approach allowed the Nuker model to adjust its inner slope to the dust-corrected values of the central pixels, but retain its outer slope shape from the initial fit. For extinctions of $A_H = 0.31, \, 0.75$, and $1.50$ mag, the value of $\gamma$ converged to values of 0.39, 0.44, and 0.47, respectively. Finally, to create dust-corrected MGE models, we replaced the $H$-band data within the disk region with the corresponding pixels in the Nuker models, and fit MGE models in \texttt{GALFIT} to these dust-corrected $H$-band images without using a mask. The major axis surface brightness profiles of these three MGE models are shown in Figure \ref{fig:dustcorrections}, and a plot of their enclosed mass profiles is shown in Figure \ref{fig:ngc1380MGEenclosedmass}. As we show in Section \ref{sec:ErrorBudget1380}, our best-fit dynamical model uses the $A_H = 0.31$ mag MGE model. We display and compare this model's isophotal contours to those of the data in Figure \ref{fig:mgecontours}. While there is good agreement between data and model outside of the central dust lane, there is some deviation between the two within the dusty region. The observed isophotes become non-elliptical towards the center, which we attribute to the presence of the dust disk. The $A_H = 0.31$ mag MGE model's components are listed in Table \ref{tab:MGE_fiducialcomponents}, while the other MGE models' components are listed in Table \ref{tab:NGC1380ExtraMGEs} in the Appendix.

\begin{deluxetable*}{c|ccc|ccc}[ht]
\tabletypesize{\small}
\tablecaption{$H$-band MGE Parameters}
\tablewidth{0pt}
\tablehead{
\multicolumn{1}{c|}{$k$} &
\colhead{$\log_{10}$ $I_{H, k}$ ($L_{\odot}\, \mathrm{pc}^{-2}$)} &
\colhead{$\sigma_{k}^{\prime}$ (arcsec)} &
\multicolumn{1}{c|}{$q_{k}^\prime{}$} & \colhead{$\log_{10}$ $I_{H, k}$ ($L_{\odot}\, \mathrm{pc}^{-2}$)} &
\colhead{$\sigma_{k}^{\prime}$ (arcsec)} & \colhead{$q_{k}^\prime{}$}\\[-1.5ex]
\multicolumn{1}{c|}{(1)} & \colhead{(2)} & \colhead{(3)} & 
\multicolumn{1}{c|}{(4)} & \colhead{(5)} & \colhead{(6)} & 
\multicolumn{1}{c}{(7)}}
\startdata
  & \multicolumn{3}{c}{\bf NGC 1380} & \multicolumn{3}{c}{\bf NGC 6861}\\
  & \multicolumn{3}{c}{$A_H=0.31$ mag} & \multicolumn{3}{c}{$A_H=0.00$ mag (Nuker Model)} \\ \cline{2-4} \cline{5-7} 
1 & 5.560  & 0.059 & 0.739  & 4.960  & 0.091  & 0.934  \\
2 & 4.996 & 0.197  & 0.710  & 4.874  & 0.179  & 0.563  \\
3 & 4.662 & 0.391  & 0.726  & 4.455  & 0.369 & 0.624 \\
4 & 4.513  &  0.743 & 0.722  & 4.175 & 0.370  & 0.624  \\
5 & 4.356 & 1.461 & 0.807 & 4.334 & 0.372  & 0.625   \\
6 & 3.973 & 3.414  & 0.613 & 4.228 & 0.630  & 0.788   \\
7 & 3.386 & 3.880  & 0.997 & 4.061 & 0.787 & 0.686   \\
8 & 3.757 & 6.018  & 0.723  & 4.391  & 1.120  & 0.512   \\
9 & 3.382 & 12.932  & 0.732  & 4.165  & 1.824 & 0.879   \\
10 & 3.043 & 18.800  & 0.400  & 4.085 & 4.227  & 0.542   \\
11 & 2.689  & 42.731  & 0.400  & 3.658 & 7.591  & 0.506  \\
12 & 2.101 & 55.077 & 0.851 & 3.299 & 13.064  & 0.562   \\
13 & 0.926 & 92.556  & 0.948 & 2.508  & 27.828 & 1.000   \\
14 & \nodata & \nodata  & \nodata  & 1.662  & 33.017 & 0.641   \\
15 & \nodata & \nodata  & \nodata & 1.684 & 52.142 & 0.883  \\
16 & \nodata & \nodata  & \nodata & 1.700  & 62.263  & 0.743  \\
\enddata
\tablecomments{NGC 1380 and NGC 6861 MGE solutions created from the combination of HST $H$-band images and best-fitting \texttt{GALFIT} Nuker models. These correspond to the $A_H = 0.31$ mag MGE model for NGC 1380 and the Nuker interpolation MGE model for NGC 6861. As described in Section \ref{sec:Results}, these two MGEs are used in the dynamical models with the lowest $\chi^2$. The first column is the component number, the second is the central surface brightness corrected for Galactic extinction and assuming an absolute solar magnitude of $M_{{\odot},H} = 3.37$ mag \citep{2018ApJS..236...47W}, the third is the Gaussian standard deviation along the major axis, and the fourth is the axial ratio, which was constrained to have a minimum value of 0.400 to allow for a broader range in the inclination angle during the deprojection process. Primes indicate projected quantities.}
\label{tab:MGE_fiducialcomponents}
\end{deluxetable*}

\subsection{NGC 6861 Host Galaxy Models}
\label{sec:NGC6861HostGalaxyModels}

\begin{figure*}
\centering
\includegraphics[width=7.0in]{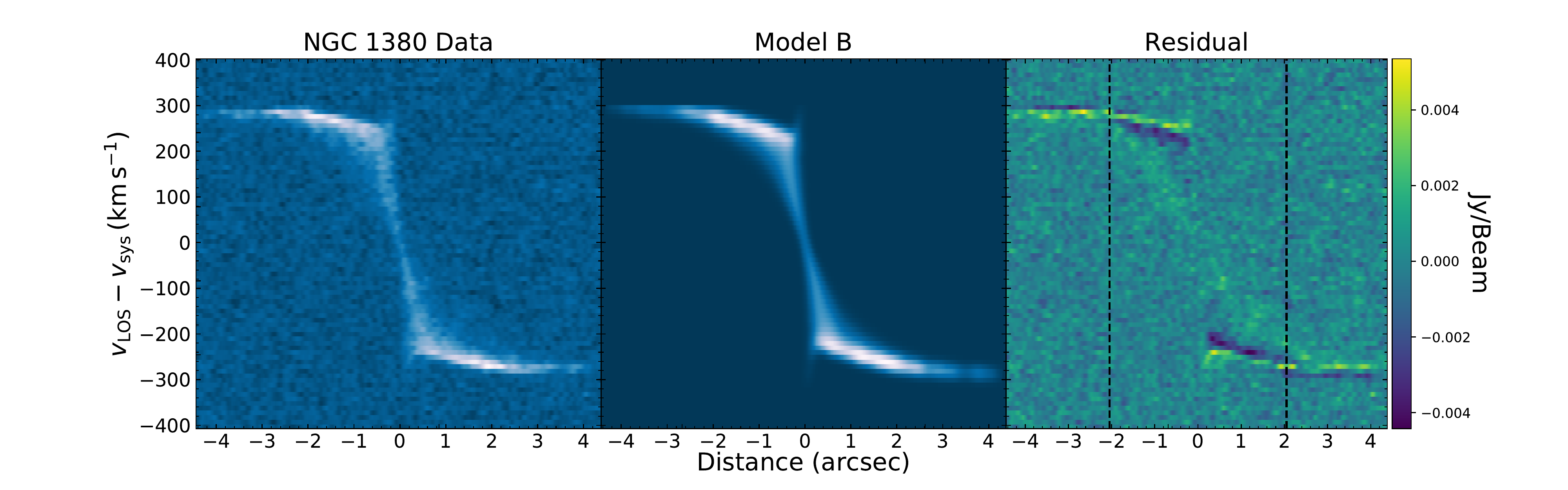}
\includegraphics[width=7.0in]{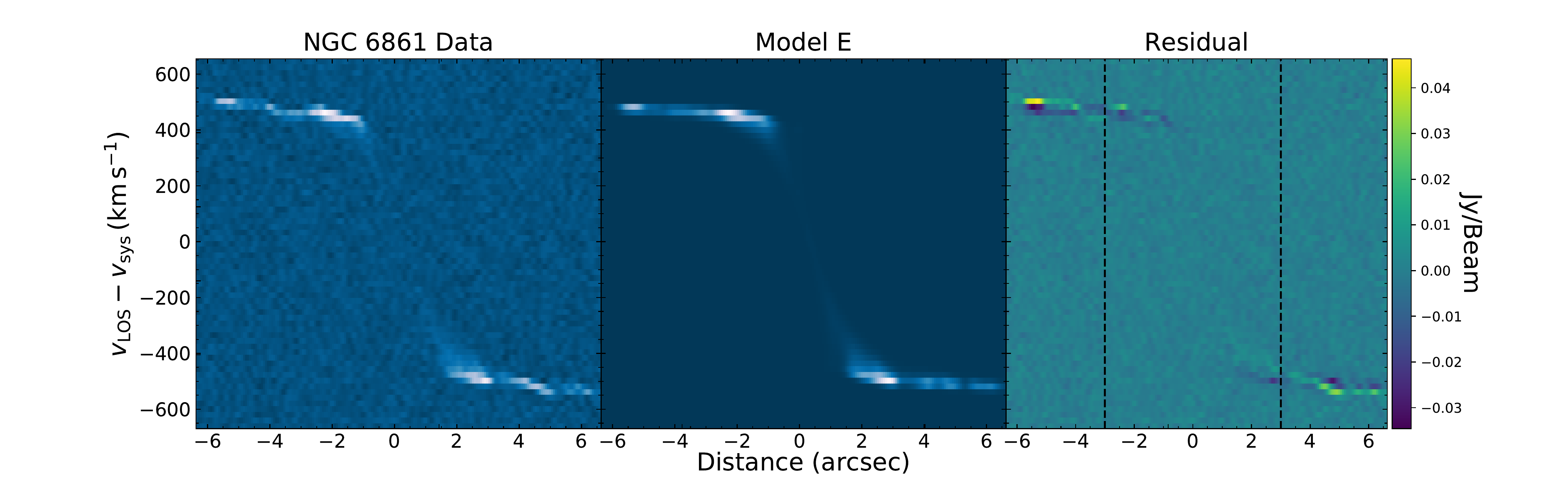}
\caption{PVDs along the major axes of both NGC 1380 (above) and NGC 6861 (below) and their respective best-fit models. Columns show ALMA Cycle 2 CO(2-1) data (left), models (center), and (data-model) residuals (right). The PVDs were generated with a spatial extraction width equivalent to a resolution element. The black dashed lines indicate the boundaries of the elliptical fit regions along the major axes.}
\label{fig:PVDs}
\end{figure*}

We created our MGE models for NGC 6861 in a different fashion from our models for NGC 1380 given the differences in their dust disks. We started again by creating a $J-H$ color map using the drizzled $J$ and $H$-band HST images of NGC 6861 to identify the pixels most affected by dust and corrected for Galactic reddening based on a foreground $A_H = 0.028$ \citep{2011ApJ...737..103S}. The color map revealed the presence of a ring-like structure within the disk with a 1$\arcsec{}$ radius hole at its center, as seen in Figure \ref{fig:jminushcolor}. A measurement of the surface brightness along the major axis of the disk (shown in Figure \ref{fig:dustcorrections}) revealed a clear decrease of stellar light due to dust. This decrease is most noticeable between $1\arcsec{}$  (the outer radius of the central hole) and $5\farcs{5}$ (the outer edge of the dust disk). The central hole is also visible in the absence of CO emission within the inner $1\arcsec{}$ of the PVD shown in Figure \ref{fig:PVDs}.

Given the lack of dynamical tracers in the central region, we wanted to test how our choice of host galaxy model impacted the inferred value of $\mbh$. We explored systematic effects due to the choice of surface brightness model used to interpolate over the dusty region, by constructing MGEs using two different models in the correction of the $H$-band image for extinction: (1) a 2D Nuker model, and (2) a 2D Core-Sérsic model \citep{2003AJ....125.2951G}. Similar to the Nuker model, the Core-Sérsic model was designed to characterize the surface brightness profiles of ETGs. Unlike the Nuker model, however, it characterizes the outer structure of ETGs with a Sérsic profile \citep{1963BAAA....6...41S} and the inner structure as a power-law. 

To fit a Nuker model to the $H$-band image, we followed a process similar to that for NGC 1380, but we did not make any adjustments to the central pixels in the drizzled $H$-band image. We first created $H$-band PSF models in Tiny Tim that were dithered and drizzled in an identical fashion to the $H$-band image, and built a mask for foreground stars, background galaxies, and the dust disk itself. Because the $J-H$ color map indicated a lack of color excess in the central hole, we masked the entire dust disk as seen in the $J-H$ image, but kept the pixels within the hole to anchor the model. We fit the inner $10\arcsec{}\times\,10\arcsec{}$ of the $H$-band image with a Nuker model in \texttt{GALFIT} and allowed all free parameters to vary. The Nuker model parameters converged to $\alpha = 0.65,\,\beta = 1.29,\,\gamma = 0.0002$, and $r_{\mathrm{b}} = 0\farcs{31}$(${\approx} 40\, \mathrm{pc}$). Finally, we replaced pixels in the original $H$-band image located in the dust disk with the corresponding pixels in the Nuker model, and proceeded to fit this image with an MGE model in \texttt{GALFIT} without using a mask. We measured and compared this MGE model's major axis surface brightness profile and isophotal contours to those of the $H$-band image in Figure \ref{fig:dustcorrections} and Figure \ref{fig:mgecontours}; we refer to this model as our Nuker interpolation model.

We constructed a 2D Core-Sérsic model for NGC 6861's $H$-band image using the \texttt{imfit} program \citep{2015ApJ...799..226E} and used the same mask, model PSF, and fitting region as for our Nuker model. The parameters that characterize the Core-Sérsic  model include the Sérsic index, $n$, break-radius, $r_{\mathrm{b}}$, effective half-light radius, $r_{\mathrm{e}}$, inner slope parameter, $\gamma$, and transition sharpness, $\alpha$. The optimization in \texttt{imfit} converged on $n=7.1,\,r_{\mathrm{b}}=3\farcs{5}\,({\approx} 455\, \mathrm{pc}), \,r_{\mathrm{e}} = 7\farcs{9}\,({\approx}1\,\mathrm{kpc}), \,\gamma = 0.61$, and $\alpha = 2.13$. We replaced the pixels in the dust disk region in the original $H$-band image with the corresponding pixels in the Core-Sérsic model, and proceeded to fit this new image with an MGE in \texttt{GALFIT}. We deprojected this MGE in an identical fashion to the MGE created with the 2D Nuker model, and refer to it as our Core-Sérsic interpolation model. We extracted its major axis surface brightness profile and compared it with both the $H$-band image and the Nuker interpolation in Figure \ref{fig:dustcorrections}. Over the extent of the dust disk, the Core-Sérsic interpolation produces higher corrected surface brightness values than the Nuker interpolation. At the nucleus, the Nuker interpolation matches the observed central surface brightness better than the Core-Sérsic interpolation, whose innermost point slightly exceeds the observed value. A comparison with the observed $H$-band isophotes is shown in Figure \ref{fig:mgecontours}. As in the case of NGC 1380, the observed isophotes became noticeably non-elliptical towards the center, although there appears to be reasonable agreement between the data and models within the central hole region. The MGE components of the Nuker interpolation are listed in Table \ref{tab:MGE_fiducialcomponents}, while the components of the Core-Sérsic interpolation are listed in Table \ref{tab:NGC6861ExtraMGEs} in the Appendix.

\section{Dynamical Modeling}
\label{sec:Dynamical Modeling}

Our dynamical modeling formalism is a Python-based adaptation of  the ALMA gas-dynamical modeling framework described by \cite{2016ApJ...822L..28B} and \cite{2019ApJ...881...10B}, which was written in the Interactive Data Language (IDL) and was used by those authors to measure the BH masses in NGC 1332 and NGC 3258. We describe the methods used in the Python version  and the modifications that differentiate it from its IDL antecedent. 

\subsection{Method}
\label{subsec:Methods}

Modeling the observed gas kinematics in an ALMA data cube relies on a few key steps and assumptions. First, we assume that the gas is distributed in a thin disk and is in circular rotation.  A model velocity field is built on a grid that is oversampled relative to an ALMA spatial pixel by a factor of $s = 3$, such that each pixel is subdivided into $s \times s = 9$ sub-pixels in order to model steep velocity gradients near the disk's center. The disk's velocity field is determined by the enclosed mass at a given radius, which consists of a central BH, the stellar mass profile of the host galaxy and a corresponding mass-to-light  ($M/L$) ratio $\Upsilon$, and the mass profile of the gas disk. For a given disk inclination $i$ and a major axis position angle $\Gamma$ (both of which are free parameters), and an assumed (fixed) distance to the galaxy, $D$, we calculate the LOS projection of this velocity field as seen on the plane of the sky. The construction and geometry of the model disk are as described by \cite{1997ApJ...489..579M} and \cite{2001ApJ...555..685B}. The LOS velocity projections are used to generate a model cube that we can compare directly to the ALMA data. For each sub-pixel with CO emission, we assume that the emergent line profile along the spectral dimension is intrinsically Gaussian. The Gaussian's line centroid and line width can be calculated at each sub-pixel by transforming both the LOS velocity projections and a spatially uniform turbulent velocity dispersion term, $\sigma(r) = \sigma_0$, into observed frequency units using the redshift $z_{\mathrm{obs}}$ (related to the systemic velocity through $v_{\mathrm{sys}} = cz_{\mathrm{obs}}$). The model cube must have its line profiles scaled by a model CO flux map, have each of its frequency slices convolved with the ALMA synthesized beam, and be downsampled to an appropriate resolution before being fitted to the ALMA data cube. We discuss these steps in further detail in the subsequent paragraphs and in Section \ref{sec:ModelOptimization}.

In total, our dynamical models use a minimum of nine free parameters: the BH mass $\mbh$, the stellar $H$-band $M/L$ ratio $\Upsilon_H$, the disk's dynamical center in pixels ($x_{\mathrm{c}},y_{\mathrm{c}}$), the disk's inclination and major axis position angle $i$ and $\Gamma$, the turbulent velocity dispersion $\sigma$, the observed redshift $z_{\mathrm{obs}}$, and a flux-scaling factor $F_0$, that correctly normalizes the model to the data. 

The circular velocity $v_c$ (relative to the disk's systemic velocity, $v_{\mathrm{sys}}$) as a function of radius is calculated as \small 
\begin{equation} v_{\mathrm{c}}(r) = \left(\frac{G\mbh}{r} + \frac{\Upsilon_H }{\Upsilon_{\mathrm{MGE}}}v^2_{\star,\,\mathrm{MGE}}(r) + v^2_{\mathrm{gas}}(r)\right)^{1/2}, \end{equation} \normalsize where $v_{\star,\,\mathrm{MGE}}$ and $v_{\mathrm{gas}}$ are the circular velocities due to the gravitational potential of the stars and the gas disk, respectively. The BH is modeled as a point mass, while the stellar and gas mass distributions are radially extended and are constructed using different methods. We modeled the stellar mass distribution using the MGE method described in Section \ref{sec:NGC1380HostGalaxyModels} and Section \ref{sec:NGC6861HostGalaxyModels}. We deprojected the MGE under the assumptions that NGC 1380 and NGC 6861 are oblate and axisymmetric and have inclination angles of $77^{\circ}$ and $73^{\circ}$, respectively, based on initial gas-dynamical modeling runs. We calculated the contribution to the circular velocity from the stars in the midplane of each disk by using the \texttt{mge\_vcirc} routine from the \texttt{JamPy} package in Python \citep{2008MNRAS.390...71C} to derive a fiducial velocity profile from our MGEs. Ideally, one should match the stellar inclination angle of the MGEs to the inclination angle found for the gas disk, as mismatches between the two lead to non-equilibrium configurations for the disk. However, this matching process is difficult to implement within our framework, and we found that the differences between the stellar and gas inclination angles were small $(<2^{\circ})$ in both NGC 1380 and NGC 6861. We will explore this aspect of the modeling process in a future work. We use $\Upsilon_{\mathrm{MGE}} = 1$ when deriving $v_{\star,\,\mathrm{MGE}}$ for each galaxy. At each model iteration, $v_{\star,\,\mathrm{MGE}}^2$ is scaled by the ratio  $\Upsilon_H/\Upsilon_{\mathrm{MGE}}$, which scales the stellar mass profile by the free parameter $\Upsilon_H$. 

As stated earlier, \cite{2017ApJ...845..170B} created mass profiles for both gas disks by averaging the CO flux in elliptical annuli centered on the continuum peaks. They determined $M_{\mathrm{gas}}$ to be $(8.4 \pm 1.6) \times 10^7 \,M_{\odot}$ and $(25.6 \pm 8.9) \times 10^7 \,M_{\odot}$ for NGC 1380 and NGC 6861, respectively, but their gas mass profiles did not assume specific shapes for the mass distributions. We assumed the mass was distributed in a thin disk and numerically integrated the projected surface mass densities to determine each gas disk's contribution to the circular velocity ($v_{\mathrm{gas}})$ using Equation 2.157 from  \cite{2008gady.book.....B}.

We disregard the mass contribution of dark matter in our models, as the stars are expected to dominate the mass budget across the length of the circumnuclear disk. For NGC 6861, we estimated the dark matter mass within the region we fit our models ($r \approx$ 400 pc) by integrating the spherical cored logarithmic density profile used by \cite{2013AJ....146...45R} in their stellar-dynamical BH mass measurement. Their model suggests an enclosed dark matter mass between $10^6-10^7 M_{\odot}$, which is lower than our estimated stellar mass by two to three orders of magnitude. 

The Gaussian line profiles at each point on the disk must be weighted by an observed CO flux map obtained from the ALMA observation. To create this flux map, we first visually identified channels in the data cube that contained CO emission. In these channels, we created a unique mask that separated pixels with visible emission from those without any. Spatially, each mask has the same size as a single frequency slice, with pixel values set to unity if the corresponding data cube pixel displays CO emission and zero if not. A channel that displays no emission would have a corresponding mask with all of its elements set to zero. An entire mask is three-dimensional, with the same dimensions as the ALMA data cube. We then multiplied each slice of this mask by the corresponding slice in the ALMA data cube, and summed the products along the spectral axis. This approach produced a less noisy image of the CO flux than if we had simply summed the data cube across channels with visible emission without any masking. To deconvolve this image, we applied five iterations of the Richardson-Lucy algorithm  \citep{1972JOSA...62...55R,1974AJ.....79..745L}. The deconvolution is performed with an elliptical Gaussian PSF that matches the specifications of the ALMA synthesized beam and uses the Richardson-Lucy algorithm  implemented in the \texttt{scikit-image} package in Python \citep{van2014scikit}. The deconvolved image is initially constructed on the original ALMA pixel scale, with each pixel then subdivided into a $s \times s$ grid of sub-pixels that matches the dimensions of the oversampled model grid. The scaled and deconvolved CO flux map is normalized so that the line profiles at each sub-pixel element for a given original ALMA pixel have equal fluxes. Thus, if the total flux in an original ALMA pixel is $F$, each sub-pixel in the $s \times s$ grid has a flux of $F/s^2$.

 The next steps consist of rescaling the oversampled model back to the original ALMA pixel scale, convolving each frequency channel within it with the ALMA synthesized beam, and minimizing $\chi^2$ between data and model. Ideally, beam convolution should occur on  the oversampled spatial grid for the highest model fidelity. However, beam convolution is the most time-consuming part of the entire modeling process and becomes prohibitively slow for oversampling factors of $s > 3$. Both \cite{2016ApJ...823...51B} and \cite{2019ApJ...881...10B} found that modeling results do not change appreciably if the convolution step is done on the original ALMA pixel scale, so we followed the same approach. We summed each $s\times s$ group of sub-pixels in our oversampled model to form a single pixel on the original ALMA scale, and then convolved each frequency slice of our model with the ALMA synthesized beam, using the \texttt{convolution} implementation in the \texttt{astropy} package for Python \citep{astropy:2013,astropy:2018}.

\subsection{Model Optimization}
\label{sec:ModelOptimization}

\begin{figure*}
    \centering
    \includegraphics[width=6.5in]{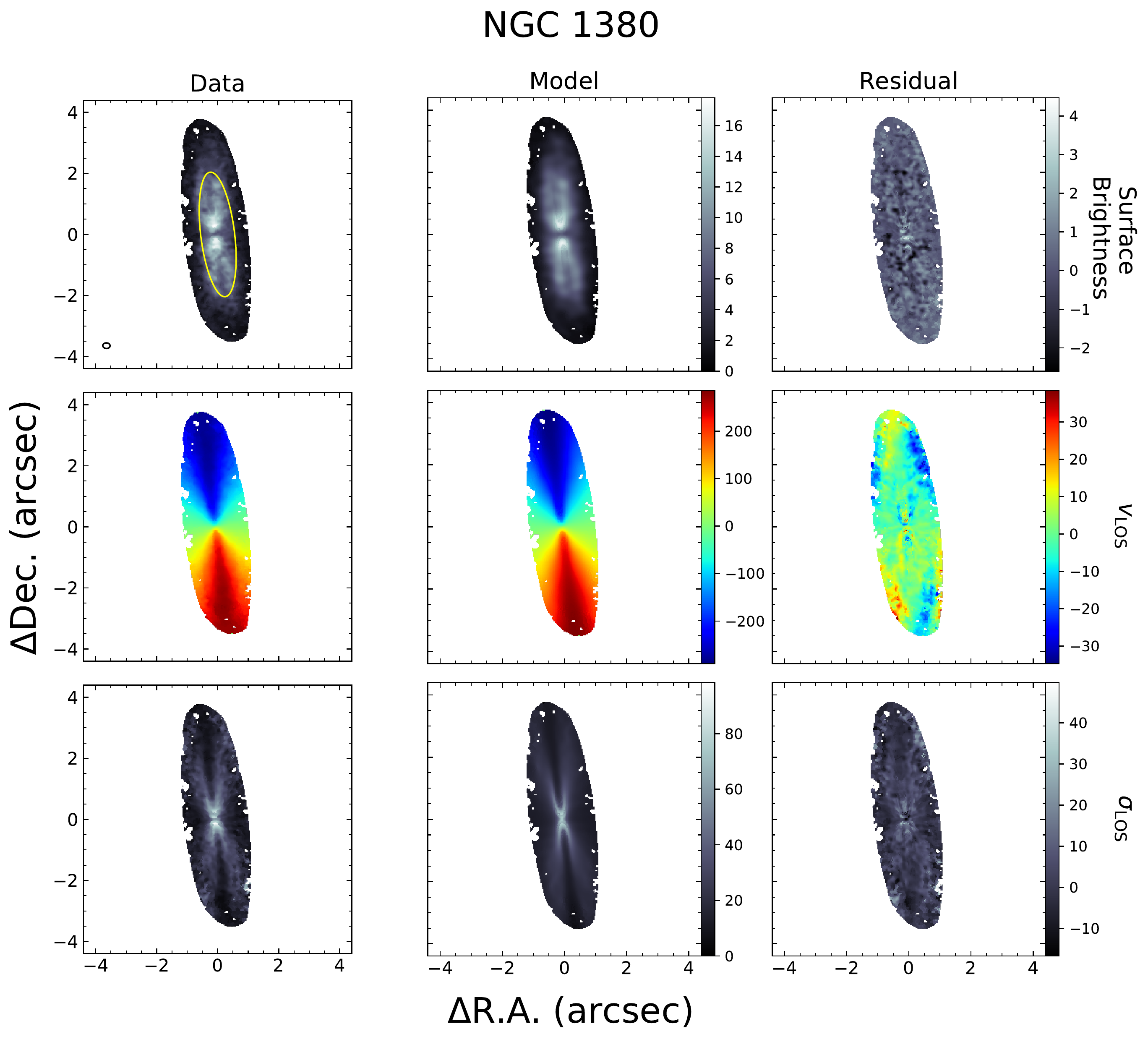}
    \caption{Moment maps for NGC 1380 constructed from the ALMA CO(2-1) data cube (left) and its fiducial model (center, model B; see Section \ref{sec:NGC1380ModelResults}). Shown are maps of moments 0, 1, and 2, corresponding to surface brightness, line-of-sight velocity $v_{\mathrm{LOS}}$, and turbulent velocity dispersion $\sigma_{\mathrm{LOS}}$. The units for the surface brightness map are mJy $\mathrm{km}\,\mathrm{s}^{-1}\,\mathrm{pixel^{-1}}$, and the units for the $v_{\mathrm{LOS}}$ and $\sigma_{\mathrm{LOS}}$ maps are $\mathrm{km}\,\mathrm{s}^{-1}$. The systemic velocity of $1854 \, \mathrm{km\, s^{-1}}$ estimated from our dynamical models has been removed from $v_{\mathrm{LOS}}$. Maps of (data-model) residuals are shown in the rightmost column.  While the line profile fits have been determined at each pixel of the full disk, the elliptical fitting region used in calculating $\chi^2$ is denoted in the top left panel with a yellow ellipse. The synthesized beam is represented by an open ellipse in the bottom left corner of the same image.}
    \label{fig:NGC1380Moments}
\end{figure*}

\begin{figure*}
    \centering
    \includegraphics[width=6.5in]{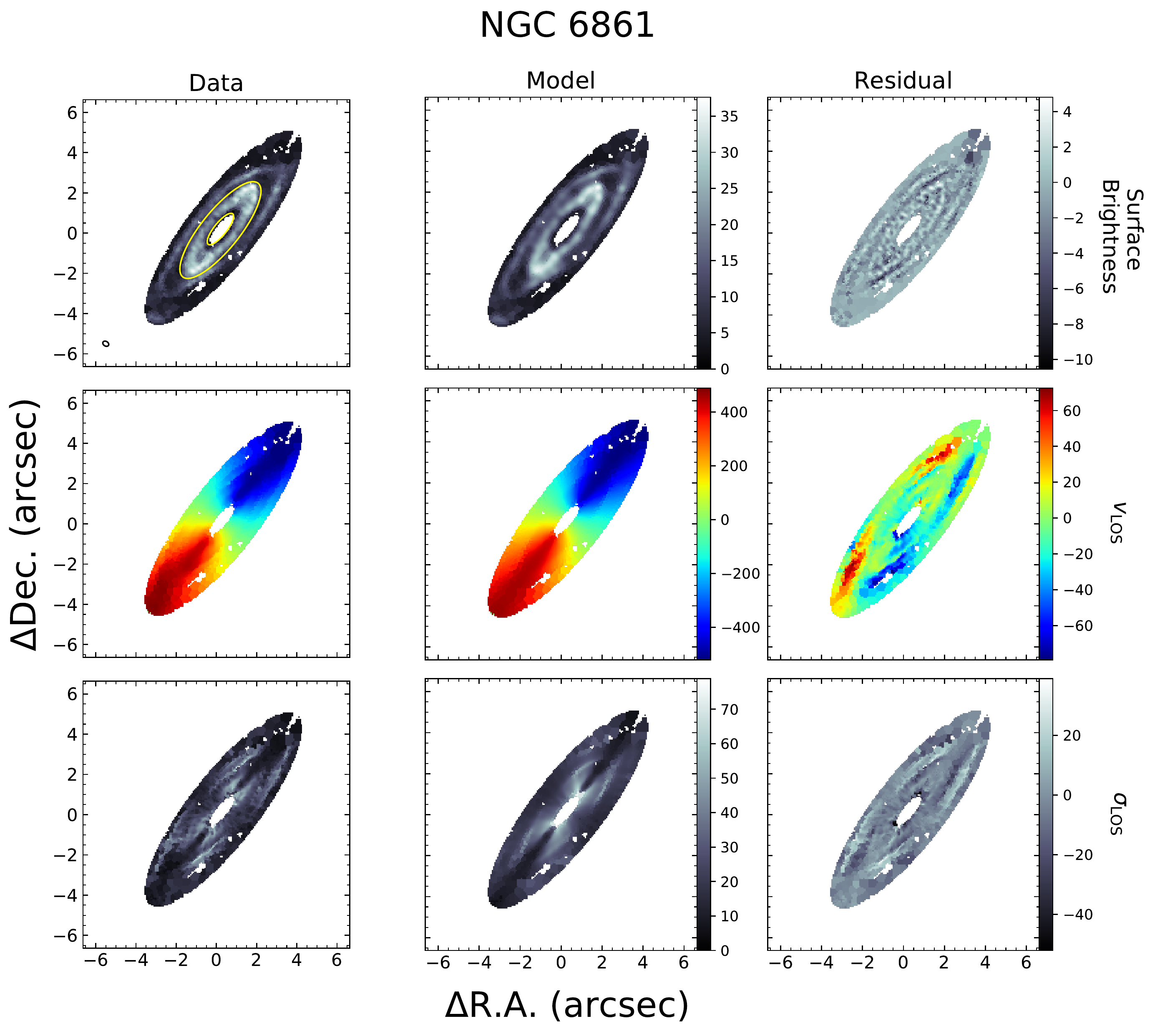}
    \caption{Moment maps for NGC 6861 constructed from the ALMA CO(2-1) data cube (left) and our model E (center). Shown are maps of moments 0, 1, and 2 corresponding to surface brightness, line-of-sight velocity $v_{\mathrm{LOS}}$, and turbulent velocity dispersion $\sigma_{\mathrm{LOS}}$. The units for the surface brightness map are mJy $\mathrm{km}\,\mathrm{s}^{-1}\,\mathrm{pixel^{-1}}$, and the units for the $v_{\mathrm{LOS}}$ and $\sigma_{\mathrm{LOS}}$ maps are $\mathrm{km}\,\mathrm{s}^{-1}$. The systemic velocity of $2796\, \mathrm{km\,s^{-1}}$ estimated from our dynamical models has been removed from $v_{\mathrm{LOS}}$. A (data-model) residual is shown in the rightmost column. While the line profile fits have been determined at each pixel of the full disk, the annular fitting region used in calculating $\chi^2$ is denoted in the top left panel with a yellow annulus. The synthesized beam is represented by an open ellipse in the bottom left corner of the same panel.}
    \label{fig:NGC6861Moments}
\end{figure*}

For a given model parameter set, we create a simulated data cube with the same spatial and spectral dimensions as the ALMA data. Therefore, our models can be fitted directly to the ALMA data cubes and can be optimized by $\chi^2$ minimization. We optimized models with the Levenberg-Marquardt algorithm \citep{Leve44,Marq63} within the \texttt{LMFIT} framework \citep{2016ascl.soft06014N} in Python and fitting to pixels that lay within the elliptical regions illustrated in the data moment 0 maps in Figures \ref{fig:NGC1380Moments} and \ref{fig:NGC6861Moments}, and within the frequency channels that span the full width of the CO emission line for each pixel. We describe the fitting regions in detail in Section \ref{sec:FitRegions}.

\subsubsection{Noise Model}
\label{sec:NoiseModel}
In order to calculate $\chi^2$ and assess the goodness-of-fit of our models to the ALMA data, we require an estimate of the flux uncertainty at each data pixel. The most straightforward approach here would be to calculate the standard deviation of pixel values in emission-line free regions of the data cube. However, the background noise in ALMA data cubes is correlated on scales comparable to the synthesized beam in each frequency channel. This correlation prevents the determination of a meaningful $\chi^2$ value without appropriate adjustments. Ideally, one would calculate a covariance matrix accounting for these correlations to compute $\chi^2$, but such an approach would be computationally expensive and challenging to implement. \cite{2016ApJ...823...51B} and \cite{2019ApJ...881...10B} adopted the simpler approach of rebinning the data by block-averaging over $m \times m$ pixel blocks within each frequency channel, where the value of $m$ was the approximate number of pixels across the width of the synthesized beam. Their method creates a data cube with a scale of approximately one rebinned pixel per synthesized beam, and mitigates the noise correlation among neighboring pixels. They then measured the standard deviation of emission-free pixels in the rebinned data cube to produce a unique value of flux uncertainty for each frequency channel, and similarly rebinned their models to compute $\chi^2$ on the block-averaged scale. 

We also incorporated the effects of the ALMA primary beam on the background noise level.  Prior to primary beam correction, the noise level in an ALMA data cube is spatially uniform, but post-correction it increases with distance from the phase center. Dynamical models are created and fitted to data cubes that have been corrected for the primary beam attenuation, so we incorporated this spatial modification into our noise model. As part of the ALMA data reduction process, a primary beam cube is generated along with the beam-corrected data cube. Multiplying the corresponding slices of these cubes together generates an uncorrected version of the data in which the background noise is spatially uniform. At this step, we block-average the data to roughly the size of the synthesized beam, using $7 \times 7$ pixel blocks for the NGC 1380 data cube and $4 \times 4$ pixel blocks for the NGC 6861 data cube. Once the data have been rebinned, we measure the standard deviations of pixel values in  blank regions of each frequency channel, and populate an array having the spatial dimensions of a block-averaged image with the value of the standard deviation at each element. To replicate the spatial modification of the noise in each channel, we block-averaged the primary beam cubes over the same pixel blocks as was done for the data and divided the block-averaged array of standard deviations by the block-averaged primary beam cube at the same frequency. In essence, we create a block-averaged noise cube that captures both the spatial and frequency dependence of the noise, which we use to compute $\chi^2$. This approach differs from previous methods where the given background noise is assumed to be spatially uniform across a given frequency slice \citep{2016ApJ...823...51B,2019ApJ...881...10B}. 

Although our noise model is designed to represent the RMS noise in emission line-free regions of each frequency channel of the data cube, an additional complication is that the mean background level can be slightly offset from zero (e.g., as a residual of imperfect passband calibration or continuum subtraction). If this is the case, the line-free regions of a cube will indirectly contribute to elevated $\chi^2$ values for model fits. For both galaxies, we find roughly equal number of channels having positive and negative mean background levels, with typical magnitudes ${\sim} 10\%$ of the respective per-channel rms noise levels. As a simple test, we empirically measured the mean background level in each frequency channel included in our fit and added this value into the corresponding channels in our synthetic model cubes. We found that our values of reduced $\chi^2$ were smaller by about ${\sim}1\%$ with this adjustment, and hence the impact of these background levels can be regarded as minimal.

\subsubsection{Fit Regions}
\label{sec:FitRegions}
We computed $\chi^2$ over elliptical fitting regions to assess the goodness of fit of our dynamical models. These elliptical fitting regions were centered on the disk centers and used the axial ratios and major axis position angles of the gas disks, as measured by \cite{2017ApJ...845..170B}. The size of the fitting region can influence the inferred value of $\mbh$. While fitting models to the entire disk uses all of the available data, the majority of pixels in the fit are at radii much greater than the BH's radius of influence. In this regime, the uncertainty in the stellar mass profile accounts for a large portion of the error budget, and model fits can lead to tight statistical constraints on both $\Upsilon_H$ and $\mbh$. If the assumed and intrinsic shapes of the stellar mass profile are discrepant, full-disk fits can force $\mbh$ to an inaccurate, but highly precise value. Alternatively, fitting to smaller regions can mitigate effects from discrepancies in the stellar mass profile because the BH mass represents a larger fraction of the total enclosed mass. Smaller fit regions can also limit systematic effects due to the structural mismatch of a thin disk model with a mildly warped disk. We discuss our selected fitting regions for NGC 1380 and NGC 6861 below.

 For NGC 1380, we initially created an ellipse centered on the disk center with an axial ratio of $q = 0.27$, and a position angle of $\Gamma = 187^{\circ}$ based on results from \cite{2017ApJ...845..170B}. For our fiducial dynamical model, we chose to fit within an ellipse that encompassed the inner half of the CO disk in order to limit the sensitivity of our dynamical models to the shape of the stellar mass profile and the disk's slightly warped structure. We also modified the size of this ellipse to see how the choice of fit region affected the inferred value of $\mbh$ in Section \ref{sec:ErrorBudget1380}. Our final fitting ellipse has a semimajor axis of $a = 2\farcs{05}$ and a semiminor axis of $b =0\farcs{55}$. This ellipse was used across 62 consecutive frequency channels that spanned the full width of the visible CO emission in the data and can be seen in Figure \ref{fig:NGC1380Moments}.  On the final rebinned scale, this choice of spatial and spectral regions resulted in 61 block-averaged pixels over 62 frequency channels for a total of 3782 data points used to calculate $\chi^2$.
 
For NGC 6861, we initially followed the same procedure, starting with the values of $q = 0.32$ and $\Gamma = 141^{\circ}$ found by \cite{2017ApJ...845..170B}. However, the disk structure in NGC 6861 is more complicated than in NGC 1380. The NGC 6861 gas disk contains a central hole that is  ${\sim}1\arcsec{}$ in radius along the major axis. Thus, the innermost CO emission is at a radius that is 3 times larger than the BH's estimated radius of influence. Additionally, the presence of rings and spiral-like substructure can be seen towards the edge of the disk. Fitting models to the entire disk led to reduced $\chi^2$ values between 2.5 and 3, as the thin disk models struggled to reproduce kinematic features in the outer disk. The inner half of the gas disk shows the most regularity in its structure, and we found that fitting dynamical models in this region led to lower reduced $\chi^2$ values and better overall fits to the data. Therefore, we created an elliptical fitting region with dimensions $a = 3\arcsec{}$ and $b = 0\farcs{96}$, as seen in Figure \ref{fig:NGC6861Moments}. In order to prevent pixels within the hole from contributing to the fit, we masked out a $1\arcsec{}$ ellipse with the same axial ratio ($q = 0.32$) at the center of our fitting region, which yielded our final annular fitting region. Along the spectral axis, we fit across 52 frequency channels that extended slightly beyond the channels with visible emission. On the final rebinned scale, with 75 rebinned pixels per channel, we included a total of 3900 data points in the fit.

\begin{figure*}
    \centering
    \includegraphics[width=7.1in]{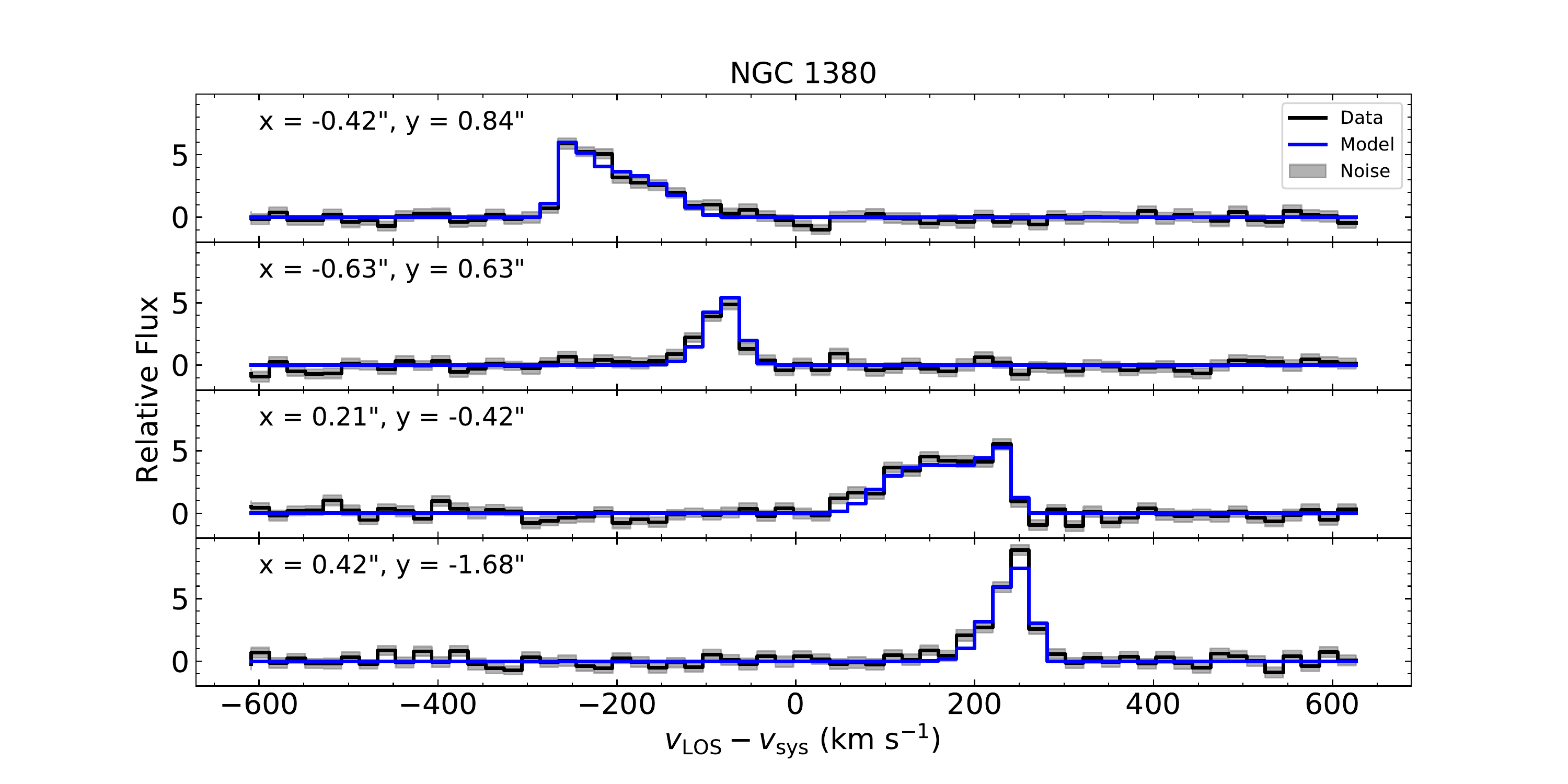}
    \includegraphics[width=7.1in]{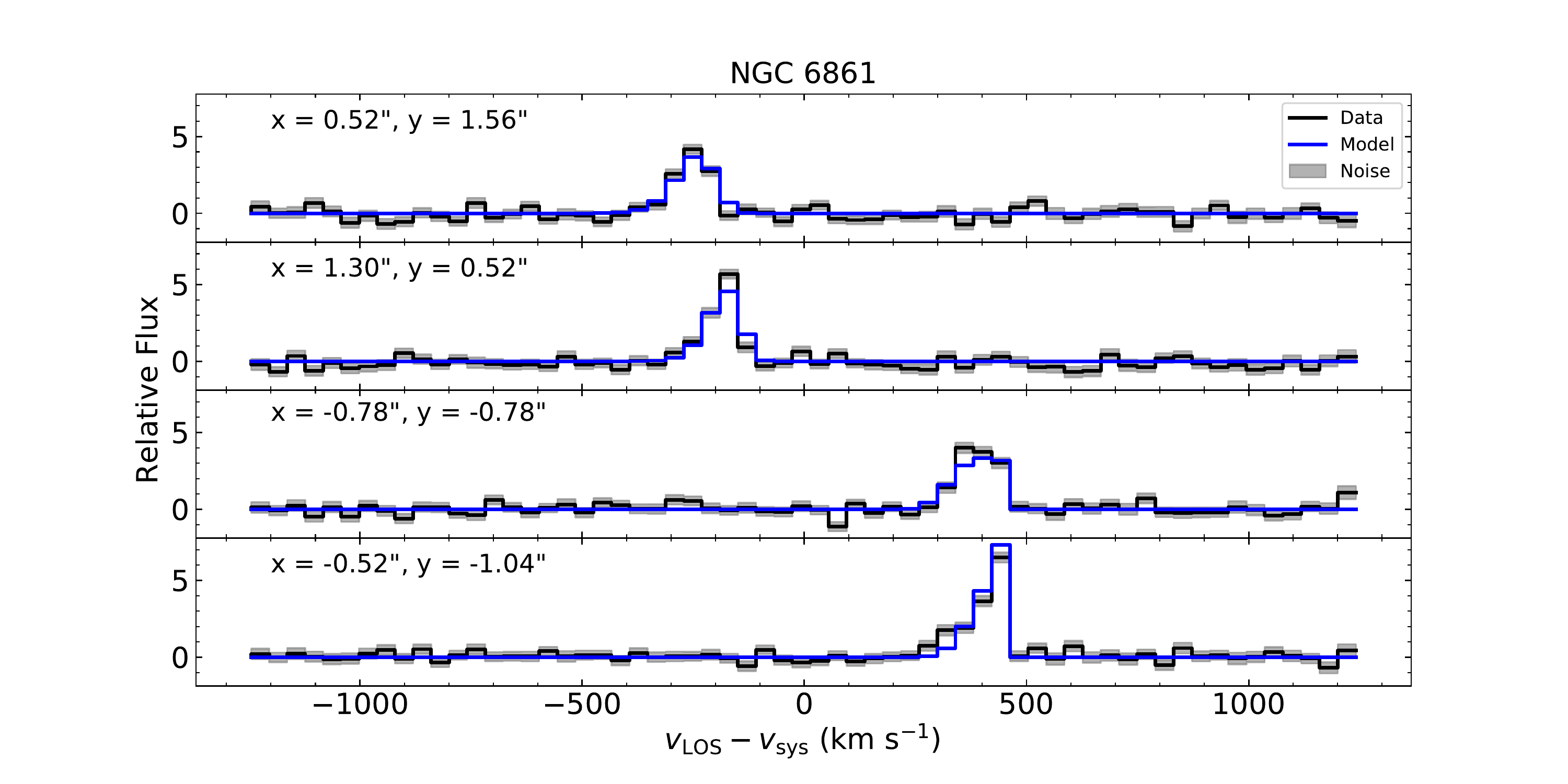}
    \caption{Representative line profiles of NGC 1380 (above) and NGC 6861 (below) measured by extracting single pixel cuts through our data, model, and noise cubes at four locations. The model line profiles are assumed to be intrinsically Gaussian at the sub-pixel level, with the asymmetric structure coming from beam-smearing. The noise band represents values in the range of $\mathrm{data} \pm 1\sigma$. These  line profiles were extracted from the cubes on the final scale of 1 block-averaged pixel per synthesized beam. The $x$ and $y$ labels indicate the pixel locations in terms of offsets in arcseconds from the disk dynamical centers, which are given in Table \ref{tabledynparams}.} 
    \label{fig:lineprofilecomparisons}
\end{figure*}

\section{Results}
\label{sec:Results}

\subsection{NGC 1380 Modeling Results}
\label{sec:NGC1380ModelResults}

We present results for four models for NGC 1380, which we refer to as models A, B, C, and D in Table \ref{tabledynparams}. The key difference among them is the input host galaxy circular velocity profile, based on one of the four MGE models described in Section \ref{sec:NGC1380HostGalaxyModels}, which accounted for four fiducial values of central dust extinction ($A_H$ = 0.00, 0.31, 0.75, and 1.50 mag for models A, B, C, and D, respectively). These dynamical models use a uniform turbulent velocity dispersion across the entire disk and are optimized over the elliptical region described in Section \ref{sec:FitRegions}. 

Models A-D yield best-fit values of $\mbh$ in the range of $(1.02 - 1.85) \times 10^8 \,M_{\odot}$, $\Upsilon_H$ between 1.30 and 1.42, and a range in reduced $\chi^2$ $(\chi^2_{\nu})$ between 1.525 and 1.563 over 3773 degrees of freedom (DOF). Our measured range of $\Upsilon_H$ is slightly above the predictions made from single stellar population (SSP) models \citep{2010MNRAS.404.1639V} that assume either a \cite{2001MNRAS.322..231K} or \cite{2003PASP..115..763C} initial-mass function (IMF) and is lower than predictions made with a \cite{1955ApJ...121..161S} IMF. These SSP models assume an old stellar population (10-14 Gyr) and solar metallicity, which are consistent with 2D IMF analyses of NGC 1380 by \cite{2019A&A...626A.124M}. As seen in Table \ref{tabledynparams}, all other free parameters remained virtually unchanged among the different dynamical models. 
We discuss the interplay between $\mbh$ and $\Upsilon_H$ in Section \ref{sec:BHSphereofInfluence}.

The major axis PVD, moment maps, and example line profiles for our best-fit model (model B) for NGC 1380 are presented and compared with the ALMA data in Figures \ref{fig:PVDs}, \ref{fig:NGC1380Moments}, and \ref{fig:lineprofilecomparisons}. The moment maps and PVD reveal that the data and model are in good agreement over a majority of the disk, although a mismatch in the observed surface brightness is seen in both the moment 0 (surface brightness) map and in the structure of the PVD, especially within the innermost ${\sim}0\farcs{5}$. Moment 1 ($v_{\mathrm{LOS}})$ maps show that data and model velocity fields are also in good agreement, although differences of ${\sim}30\,\mathrm{km}\,\mathrm{s^{-1}}$ are noticeable towards the disk edge, outside the fit region, and at the disk center, where the impact of beam-smearing is most severe. Most likely, these differences are a result of the differences between the model and intrinsic stellar velocity profiles.  The extracted line profiles of the block-averaged data and best-fit model highlight our models' ability to reproduce the observed shapes of the line profiles, even when they display asymmetric structure.

To determine the statistical uncertainties of the free parameters, we performed a Monte Carlo simulation for model B. We created 150 realizations of the best-fit model by adding noise to each pixel of the model cube. The value of the noise at each pixel was determined by choosing a random value drawn from a Gaussian distribution with a standard deviation equal to the value of the corresponding pixel in the noise cube described in Section \ref{sec:NoiseModel}. We re-fit to each noise-added model realization using the values found in Table \ref{tabledynparams} as initial guesses; the standard deviation of each recovered parameter was identified as the $1 \sigma$ uncertainty. For $\mbh$, we found a tight distribution centered at our initial guess of $\mbh = 1.47 \times 10^8 \,M_{\odot}$ with a standard deviation of $2 \times 10^6 \,M_{\odot}$, or 1.4\% of the mean. The statistical uncertainties for the free parameters are listed under model B's best-fit values in Table  \ref{tabledynparams}. Based on other Monte Carlo simulations we ran, these statistical uncertainties are representative of those for models A, C, and D.

\subsection{Error Budget for NGC 1380}
\label{sec:ErrorBudget1380}

While the statistical uncertainties from our Monte Carlo simulation are small, there are several other sources of uncertainty that stem from the choices we made when building our dynamical models. Thus, we conducted numerous tests to determine the impact these choices had on the value of $M_{\mathrm{BH}}$.

\textit{Dust Extinction}: The value of $\mbh$ in our model optimizations is highly sensitive to the choice of MGE model. Using the initial, dust-masked MGE, our models converge on $\mbh = 1.85 \times 10^8 \,M_{\odot}$. As we increase the central extinction from $A_H = 0.00$ mag to $A_H = 0.31$ mag, corresponding to a loss of 25\% of the total stellar light behind the dusty disk, $\mbh$ decreased to $1.47 \times 10^8 \,M_{\odot}$, representing a ${\sim}20\%$ decrease from the initial fit. This particular model also shows a decrease in the resultant $\chi^2_{\mathrm{\nu}}$, as model A with the initial MGE returns $\chi^2_{\mathrm{\nu}} = 1.544$, while model B with $A_H = 0.31$ mag yields $\chi^2_{\mathrm{\nu}} = 1.525$. Increasing the extinction further to $A_H = 0.75$ mag and $A_H = 1.50$ mag further decreases $\mbh$, as models C and D converge to values of $1.27 \times 10^8 \,M_{\odot}$ and $1.02 \times 10^8 \,M_{\odot}$, i.e., ${\sim}$31\% and ${\sim}$45\% decreases from the model A value, respectively. However, both models C and D result in higher $\chi^2_{\mathrm{\nu}}$ values. 

We chose the $A_H = 0.31$ mag MGE model (model B) to use as our fiducial model for a number of reasons. First, this MGE model accounts for the impact of dust extinction, whereas the initial MGE simply masks dust out. Next, the dynamical model that uses the $A_H = 0.31$ mag MGE has the lowest value of $\chi^2$ out of all our models, signifying the best overall match to the ALMA data. Lastly, extracting the major axis surface brightness profile from this MGE model reveals that it most closely resembles our surface brightness profile after correction for the low-extinction branch of the reddening curve shown in Figure \ref{fig:dustcorrections}. Therefore, for all of the remaining systematic tests, we use the $A_H = 0.31$ mag MGE as our host galaxy model.
 
\textit{Radial motion}: Although there is no evidence of strong deviations from circular motion in the NGC 1380 gas disk, we constructed a simple model that allows for radial motion in our dynamical models. We followed an approach similar to \cite{2019ApJ...881...10B} and \cite{2021arXiv210407779C} and added a radially inward velocity term to our dynamical models. We included an additional free parameter, $\alpha$, which lies in the range $[0,1]$ and controls the balance between pure rotational ($\alpha = 1)$ and radially inflowing $(\alpha = 0)$ motion. Mathematically, we defined 
$\alpha$ to be the ratio between the rotational velocity, $v_{\mathrm{rot}}$, and the ideal circular velocity in our model grid (i.e., $\alpha = v_{\mathrm{rot}}/v_{c}$). We defined the relationship between $\alpha$, radial inflow velocity, and the ideal circular velocity as $v_{\mathrm{infl}} = \sqrt{2(1-\alpha^2)} v_c$. Thus, when $\alpha = 1$, our model velocities are circular, and when $\alpha = 0$, the velocities are radially inflowing at the ideal free-fall speed of a test particle falling in from infinity. We projected the radial velocity component along the LOS, and added it linearly to the projected LOS rotation velocity at each pixel in our model grid. Upon optimizing, we found that the models strongly favored pure rotation, with a best-fitting value of $\alpha = 1$. This model converged on $\mbh = 1.47 \times 10^8 \,M_{\odot}$, leaving the results found for the fiducial model unchanged.

\textit{Turbulent velocity dispersion}: The ratio of the turbulent velocity dispersion to the rotational velocity $(\sigma/v_{\mathrm{rot}})$ determines if the disk can be treated as dynamically cold (where $(\sigma/v_{\mathrm{rot}})^2 \ll 1)$, or if dynamical pressure effects from turbulence must be accounted for.  By using our $A_H = 0.31$ mag MGE model, our radial gas mass profile, the fiducial BH mass, and our best-fit value of $\sigma_0 = 10.5\, \mathrm{km}\, \mathrm{s}^{-1}$, we find a maximum value of $\sigma_{0}/v_\mathrm{c} = 0.05$ (where we have set $v_{\mathrm{rot}} = v_{\mathrm{c}}$, the ideal circular velocity) at about $r = 50$ pc, indicating that treating the disk as dynamically cold and neglecting dynamical pressure effects are justified. Nevertheless, a spatially uniform turbulent velocity dispersion term might be insufficient to characterize possible variations in turbulence across the entire disk. Therefore, in addition to using a spatially uniform gas turbulent velocity dispersion term, $\sigma(r) = \sigma_0$, we also tried incorporating a Gaussian turbulent velocity dispersion profile $\sigma(r) = \sigma_0 + \sigma_1 \exp[-(r-r_0)^2/2\mu^2]$ into our fiducial model. This profile adds three free parameters and allows for more flexibility in characterizing the overall velocity dispersion. However, the model is not physically motivated, and is used here as a simple tool for exploring possible variations in $\sigma(r)$. The preferred turbulent velocity dispersion parameters of $\sigma_0$ = 10.6 $\mathrm{km\,s^{-1}}$, $\sigma_1$ = 0.21  $\mathrm{km\,s^{-1}}$, $r_0 = 0.06$\,$\mathrm{pc}$, and $\mu = 0.02\,\mathrm{pc}$ yield  a turbulent velocity dispersion profile that is dominated by the spatially uniform term of $10.6 \,\mathrm{km\,s^{-1}}$, and is nearly identical to the fiducial model's spatially uniform $ \sigma_0 = 10.5\, \mathrm{km\,s^{-1}}$. The modified model yields $\mbh = 1.47 \times 10^8 \,M_{\odot}$ and $\chi^2_{\nu} = 1.526$, demonstrating almost no effect on the results of the fiducial model. 

\textit{Fit region}: We tested the sensitivity of $\mbh$ to the fit regions of the dynamical models by making two separate adjustments to the spatial fitting ellipse described in Section \ref{sec:FitRegions} and used to calculate $\chi^2$. The first adjustment was expanding the fitting ellipse to cover the entirety of the gas disk, with semimajor axis length $4\farcs{1}$ and  semiminor axis length $1\arcsec{}$. For this fitting ellipse, the models converged on $\mbh = 1.63 \times 10^8 \,M_{\odot}$, an increase of $10.9\%$ from the fiducial model, and $\chi^2_{\mathrm{\nu}} = 1.451$. This larger fitting ellipse includes nearly four times as many data points, but a majority of points are at radii where the extended stellar mass distribution dominates the total enclosed mass.

Our second adjustment was reducing the fitting ellipse to fit only the inner third of the disk, with semimajor and semiminor axis lengths of $1\farcs{37}$ and $0\farcs{33}$. The resultant value of $\mbh$ was $1.52 \times 10^8 \,M_{\odot}$, an increase of $3.4\%$ relative to our fiducial model, with $\chi^2_{\mathrm{\nu}} = 1.337$. On this scale, the fit contains a higher fraction of data points that display unresolved gas kinematics, particularly along the disk's minor axis, where beam-smearing effects are more severe.

\textit{Pixel oversampling}: Other molecular gas-dynamical studies such as \cite{2016ApJ...822L..28B} have found that $\mbh$ is relatively insensitive to the choice of pixel oversampling factor, $s$. We tested our own dynamical models with oversampling factors of $s=1$ and $s=4$. For $s=1$, the result was $\mbh = 1.45 \times 10^8 \,M_{\odot}$, a decrease of 1.4\% relative to model B, with a higher $\chi^2_{\nu}$ of 1.543, as expected for no pixel oversampling. At $s=4$, the resulting $\mbh$ was $1.47 \times 10^8 \,M_{\odot}$, identical to our fiducial model result, with a slightly improved $\chi^2_{\nu}$ of 1.524. These results show that $\mbh$ has little sensitivity to the choice of $s$, even in the no-oversampling case of $s=1$.

\textit{Gas mass}: We performed two tests to observe the dependence of $\mbh$ on the inclusion or exclusion of the gas disk's contribution to the mass model. First, we optimized a dynamical model that did not include the circular velocity contribution from the gas disk which was derived from the mass surface densities in \cite{2017ApJ...845..170B}, but otherwise used the same inputs as model B. Without this contribution, the model converged upon $\mbh = 1.43 \times 10^8 \,M_{\odot}$, a decrease of 2.7\%, and the same $\chi^2_{\nu} = 1.525$ as our fiducial model. In addition, we rescaled the gas mass surface densities to the lower value implied by a CO(2--1)/CO(1--0) intensity ratio $\approx 0.9$ (see Section 2.2.2) and reoptimized our dynamical model with this adjusted circular velocity contribution. With this adjustment, our model converged upon $\mbh = 1.45 \times 10^8 \,M_{\odot}$, and an identical $\chi^2_{\nu} = 1.525$. These results suggests the inclusion or exclusion of the gas component in models can be important in percent-level precision BH mass measurements. However, for our measurements, it is a relatively minor contribution to the error budget in comparison to that of the dust extinction. 

\textit{Unresolved Active Galactic Nucleus Emission}: Given that there is evidence of a weak AGN in NGC 1380, we explored the possibility that our MGEs could be incorporating the light from this AGN in addition to the stars. As a test, we removed the innermost component (FWHM = $0\farcs{14}$ = 11.5 pc)  of the $A_H = 0.31$ mag MGE model, and deprojected the remaining components. Using this altered MGE model, the BH mass rose to $\mbh = 2.04 \times 10^8 \, M_{\odot}$, an increase of 38.7\% from the fiducial BH mass ($1.47 \times 10^8\, M_{\odot}$) found when we included the innermost component, but the reduced $\chi^2$ value also rose to 1.528, indicating a poorer fit to the data. If we deproject this innermost component individually, and assume it is composed entirely of starlight, the corresponding stellar mass is $0.59 \times 10^8\, M_{\odot}$ (assuming $\Upsilon_H = 1.42$), which is slightly higher (by $2 \times 10^6\, M_{\odot}$) than the difference in BH mass. While it is difficult to determine the amount of AGN light in the innermost MGE component, given that the increase in BH mass between the two dynamical models is commensurate with the decrease in assumed stellar masses, this test shows that the BH and stellar mass of the innermost component are degenerate.

The large range of $M_{\mathrm{BH}}$ found in Table \ref{tabledynparams} and the sensitivity tests performed above show that the systematic error is dominated by the uncertainties associated with the host galaxy models. Specifically, these uncertainties are associated with the amount of stellar mass, which changes depending on the assumed dust extinction and/or the presence or absence of an AGN contributing to the central host galaxy light, as shown when we removed the innermost component of our MGE. We chose model B as our fiducial model for the reasons described in Section \ref{sec:ErrorBudget1380} and treat the $\Delta M_{\mathrm{BH}}$ between its $\mbh$ value of $1.47 \times 10^8 \,M_{\odot}$ and the maximum ($1.85 \times 10^8 \,M_{\odot}$) and minimum $M_{\mathrm{BH}}$ ($1.02 \times 10^8 \,M_{\odot}$) values found in Table \ref{tabledynparams} as a rough estimate of the uncertainty due to the dust correction. These maximum and minimum values are about 26\% larger and 31\% smaller than the fiducial $\mbh$ value, respectively. 

It is clear that the systematic uncertainties exceed the statistical uncertainty (${\approx 1\%}$) and the uncertainty associated with the distance to the galaxy (${\approx 8\%}$) \citep{2001ApJ...546..681T}. Considering that the uncertainties from the dust correction and the host galaxy mass modeling dominate the total systematic uncertainty, we adopt the BH mass of $1.47 \times 10^8\, M_{\odot}$ from our fiducial model and the aforementioned uncertainties as our estimate for $\mbh$. Therefore, the expected range of $\mbh$ in NGC 1380 is $(1.02 - 2.04) \times 10^8 \, M_{\odot}$.

\begin{deluxetable*}{ccccccccccccc}[ht]
\tabletypesize{\small}
\tablecaption{Dynamical Modeling Results}
\tablewidth{0pt}
\tablehead{
\colhead{Model} & 
\colhead{MGE} &
\colhead{\mbh} & 
\colhead{$\Upsilon_H$} & 
\colhead{$i$} &
\colhead{$\Gamma$} &
\colhead{$\sigma_0$} &
\colhead{$x_{\mathrm{c}}$} &
\colhead{$y_{\mathrm{c}}$} & 
\colhead{$v_{\mathrm{sys}}$} &
\colhead{$F_0$} & 
\colhead{$\chi^2_{\nu}$}
\\[-1.5ex]
\colhead{} & 
\colhead{} &
\colhead{} & 
\colhead{($M_{\odot}/L_{\odot}$)} & 
\colhead{($^{\circ}$)} &
\colhead{($^{\circ}$)} &
\colhead{$(\mathrm{km}\,\mathrm{s^{-1}})$} &
\colhead{(arcsec)} &
\colhead{(arcsec)} &
\colhead{$(\mathrm{km}\,\mathrm{s^{-1}})$} &
\colhead{}
} 
\startdata
\hline \multicolumn{11}{c}{\textbf{NGC 1380}} \\
& ($A_H$ mag) & ($10^8 \,M_{\odot}$) & & & & & & & & & & \\
A & 0.00 & 1.85 & 1.33 & 76.9 & 187.1 & 10.8 & 0.016 & 0.013 & 1853.83 & 0.99 & 1.544 \\
\textbf{B} & 0.31 & 1.47 & 1.42 & 76.9 & 187.2 & 10.5 & 0.017 & 0.013 & 1853.86 & 0.99 & 1.525\\
& & (0.02) & (0.003) & (0.04) & (0.04) & (0.13) & (0.001) & (0.001) & (0.15) & (0.003) & \\
C & 0.75 & 1.27 & 1.36 & 76.9 & 187.2 & 10.5 & 0.017 & 0.013 & 1853.86 & 0.99 & 1.545 \\
D & 1.50 & 1.02 & 1.30 & 76.8 & 187.2 & 10.5 & 0.017 & 0.013 & 1853.88 & 0.99 & 1.563 \\
\hline \multicolumn{11}{c}{\textbf{NGC 6861}} \\
& (Nuclear Profile) & ($10^9 \,M_{\odot}$) & & & & & & & & & \\ 
\textbf{E} & Nuker &  1.13 & 2.52 & 72.7 & 142.5 & 7.2 & 0.057 & 0.038 & 2795.63 & 1.03 & 1.987 \\
& & (0.04) & (0.01) & (0.07) & (0.05) & (0.29) & (0.002) & (0.002) & (0.29) & (0.006) & \\
F & Core-Sérsic & 2.89 & 2.14 & 73.6 & 142.6 & 7.4 & 0.049 & 0.046 & 2795.65 & 1.04 & 2.004  
\enddata
\begin{singlespace}
\tablecomments{Best-fit parameter values obtained by fitting thin disk dynamical models to the NGC 1380 and NGC 6861 CO(2-1) data cubes. We derive 1$\sigma$ statistical uncertainties for the parameters of fiducial models B and E, based on a Monte Carlo resampling procedure described in Section \ref{sec:NGC1380ModelResults} and list them under the results for models B and E. These models have 3773 and 3891 degrees of freedom, respectively.  The major axis PA, $\Gamma$, is measured east of north for the receding side of the disk. The disk dynamical center, $(x_\mathrm{c}, y_{\mathrm{c}})$ is measured in arcsecond offsets from the nuclear continuum centroids for NGC 1380 and NGC 6861 determined in \cite{2017ApJ...845..170B}. The observed redshift, $z_{\mathrm{obs}}$, is used in our dynamical models as a proxy for the systemic velocity of the disk, $v_{\mathrm{sys}}$, in the barycentric frame via the relation: $v_{\mathrm{sys}} = cz_{\mathrm{obs}}$ and is used to translate the model velocities to observed frequency units.}
\end{singlespace}
\label{tabledynparams}
\end{deluxetable*}

\subsection{NGC 6861 Modeling Results}
\label{sec:NGC6861DynModResults}

We optimized two different dynamical models for NGC 6861, which we refer to as models E and F in Table \ref{tabledynparams}. The difference between them is the input stellar circular velocity profile, based on one of the two NGC 6861 MGE models described in Section \ref{sec:NGC6861HostGalaxyModels} which model the nuclear region in the $H$-band image with either a Nuker (E) or Core-Sérsic (F) model. These two dynamical models used a uniform turbulent velocity dispersion across the entire disk and are optimized over the annular region described in Section \ref{sec:FitRegions}. 

Model E converges on $\mbh = 1.13 \times 10^9 \,M_{\odot}$ and $\Upsilon_H = 2.52$ with $\chi^2_{\nu} =  1.987$, while model F returns values of $\mbh = 2.89 \times 10^9 \,M_{\odot}$ and $\Upsilon_H = 2.14$ with $\chi^2_{\nu} = 2.004$. The $\Upsilon_H$ values are higher than the ranges predicted by the SSP models of \cite{2010MNRAS.404.1639V}. Similarly, the range of $\Upsilon_I = (5.7-6.3)$ determined observationally by \cite{2013AJ....146...45R} is higher than SSP model predictions.  All other free parameters remain consistent between the two models, although the inclination angle $i$ slightly increases from $72.7^{\circ}$ to $73.6^{\circ}$ from model E to F. 

We created moment maps and a major axis PVD, and extracted line profiles for model E to compare with the ALMA data. The residuals between the data and model $v_{\mathrm{LOS}}$ maps show that our thin disk model emulates the data's observed $v_{\mathrm{LOS}}$ well within our designated fitting region, but discrepancies in excess of ${\sim} 60\, \mathrm{km}\,\mathrm{s^{-1}} $ are seen at larger radii. These discrepancies highlight kinematic substructure within the disk at these larger radii that our models are unable to reproduce, although given that the maximum value of $\sigma_0/v_{\mathrm{c}}$ across the NGC 6861 disk is ${\sim} 0.02$, our treatment of the disk as dynamically cold is justified. The PVD also shows discrepancies along the major axis, as structural differences between the data and model PVDs are prevalent at radii larger than ${\sim}3\arcsec{}$, which corresponds to the semimajor axis of our elliptical fitting region. The extracted line profiles in Figure \ref{fig:lineprofilecomparisons} show that our models are able to reproduce the observed line profile shapes well, although slight inconsistencies in the peak amplitude, in terms of both overall height and velocity channel, are evident in some spectra. 

We conducted Monte Carlo simulations to estimate fitting uncertainties for both model E and F as we did for model B. We generated 150 realizations of each model cube by adding noise to each pixel as described in Section \ref{sec:NGC1380ModelResults}. For model E, we found a distribution centered at its best-fit $\mbh$ value of $1.13 \times 10^9 \,M_{\odot}$ with a standard deviation of $4 \times 10^7 \,M_{\odot}$, or 1.4\% of $\mbh$. Model F's Monte Carlo simulation was centered at $\mbh = 2.89 \times 10^9 \,M_{\odot}$ and also had a standard deviation of $4 \times 10^7 \,M_{\odot}$, or 3.5\% of $\mbh$. We chose to use the standard deviations of each of the free parameters from model E as representative of statistical uncertainties associated with these values and list them in Table \ref{tabledynparams}.

The Monte Carlo simulations show that the statistical model-fitting uncertainties are significantly smaller than the systematic uncertainty associated with our choice of host galaxy model, which return values of $\mbh$ that are different by a factor of ${\sim}3$. Because of this large difference in $\mbh$ between models E and F, we did not perform extensive systematic tests on these models as we did for model B in Section \ref{sec:ErrorBudget1380}, as the uncertainty associated with our choice of host galaxy model dominates the total error budget.

To determine a lower limit on $\mbh$, we adjusted the central flux in NGC 6861's $H$-band image in the same manner as was done for NGC 1380's $H$-band image in Section \ref{sec:NGC1380HostGalaxyModels}. We corrected our Nuker interpolation MGE model under the assumption that the disk resides in the midplane of the galaxy and that only the starlight originating from behind the dust disk experiences any extinction. Our modified Nuker interpolation model explored the extreme limit where $A_H = \infty$, signifying that all of the light behind the disk is lost, and raising the innermost value of the major axis surface brightness profile by 0.75 mag\,$\mathrm{arcsec}^{-2}$. Even with this maximally peaked surface brightness model, the value of $\mbh$ was non-zero and converged on $9.7 \times 10^7\, M_{\odot}$, which serves as our measurement's lower limit. We emphasize that even with the assumption that the central region of NGC 6861 is optically thick (which is highly unlikely given the $J-H$ color map) and the absence of dynamical information within the inner $1\arcsec{}$, our dynamical models still require a central compact mass to reproduce the observations.

\section{Discussion} 
\label{sec:Discussion}

Our molecular gas-dynamical measurements are the first and second attempts to determine the masses of the central BHs in  NGC 1380 and NGC 6861, respectively. For each galaxy, the presence of dust limits the measurement precision on $\mbh$. In NGC 1380, we find $\mbh = 1.47 \times 10^8$ with an uncertainty of ${\sim} 40\%$ which is dominated largely by the dust corrections. The measurement precision for NGC 6861's BH is even more limited due to the lack of dynamical tracers within the BH's sphere of influence, as the resulting values of $\mbh$ differ by a factor of ${\sim}3$ depending on the model used for the host galaxy. Below, we discuss the importance of resolving the BH's sphere of influence and  accounting for the presence of dust in both galaxies. We also discuss how parameter degeneracies emerge within our models from these factors, and we compare our measured $\mbh$ values to predictions from the BH-host galaxy scaling relations given by \cite{2013ARAA..51..511K}.

\subsection{The BH Sphere of Influence}
\label{sec:BHSphereofInfluence}

\begin{figure*}[t]
    \centering
    \includegraphics[scale=0.47]{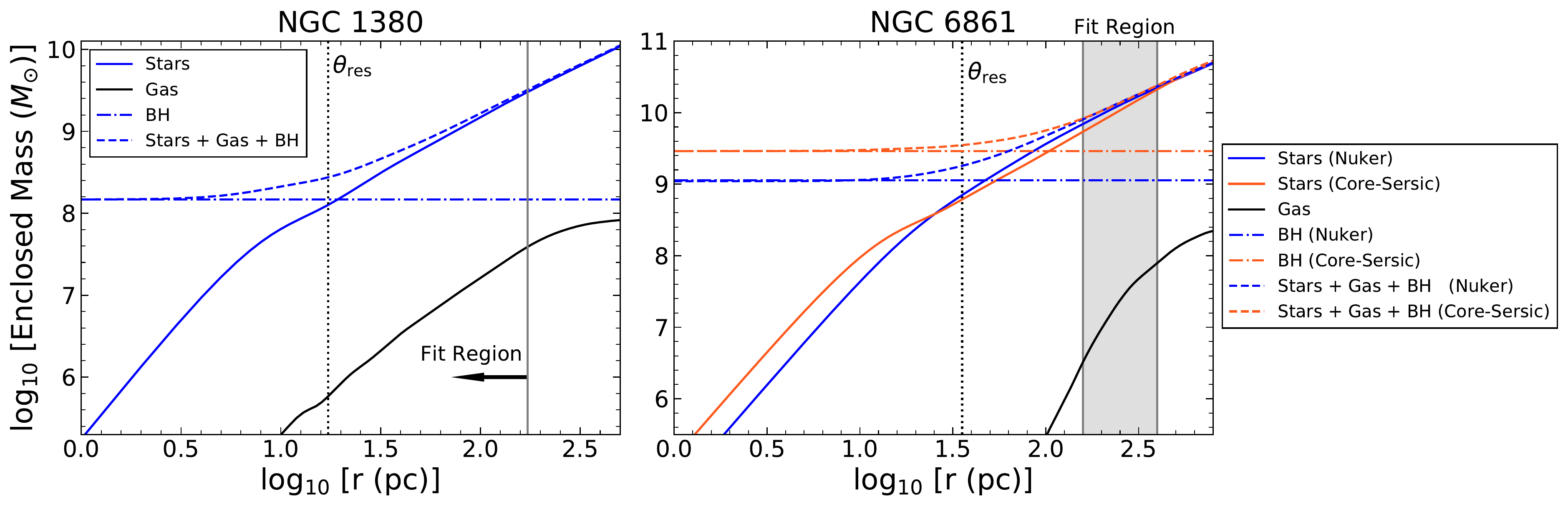}
    \caption{Enclosed mass profiles as a function of radius used in the dynamical modeling of NGC 1380 (left) and NGC 6861 (right). We show the mass contributions from the stars, gas, and BH, as well as the sums of their masses. For NGC 1380, we display the stellar mass profile associated with the $A_H = 0.31$ mag MGE used in our fiducial model. For NGC 6861, the enclosed mass profiles associated with our Nuker interpolation MGE model fit are shown in blue, while those associated with our Core-Sérsic interpolation MGE model are shown in orange. The gray shaded region indicates the radii over which dynamical modeling fits were performed along the major axis. The dotted lines correspond to the 0\farcs{21} and  0\farcs{28} resolutions of the NGC 1380 and NGC 6861 Cycle 2 observations, respectively.}
    \label{fig:mass_coresersic_nuker}
\end{figure*}

 The precision of a BH mass measurement is highly dependent on how well the observations resolve the radius of the BH's dynamical sphere of influence, $r_g$. Within $r_g$, the central BH is the dominant contributor to the total gravitational potential. If $r_g$ is unresolved, then the measured value of $\mbh$ depends heavily on the accuracy of the assumed host galaxy model, and dynamical models are susceptible to parameter degeneracies. For each galaxy, we estimated $r_g$ in two distinct  ways. The first was to determine the radius where $\mbh$ is equal to the enclosed stellar mass, and the second was to calculate $r_g \approx G\mbh/\sigma_{\star}^2$ using our measured values of $\mbh$ and known literature values of $\sigma_{\star}$ for both galaxies. 

For NGC 1380, we used the best-fit value of $\mbh = 1.47 \times 10^8 M_{\odot}$ from our fiducial model to determine $r_g$. The radius where the enclosed stellar mass (derived from the MGE used in the fiducial model) equaled $\mbh$ was $r_g$ = 18 pc ($0\farcs{22}$), which is nearly identical to our average beam size of $0\farcs{21}$. If we instead calculate $r_g$ as $r_g = G\mbh/\sigma_{\star}^2$, using the average value of $\sigma_{\star} = 215\,\mathrm{km}\,\mathrm{s^{-1}}$ from Hyperleda \citep{2014A&A...570A..13M}, we find $r_g = 14$ pc ($0\farcs{17}$). We display both the total enclosed mass profile and the separate contributions from each component to our dynamical model in Figure \ref{fig:mass_coresersic_nuker}. These estimates demonstrate it is likely that the observations only marginally resolve $r_g$. 

While there is evidence of a slight central upturn in gas velocity within the innermost ${\sim} 0\farcs{2}$ of NGC 1380, given that the observations do not fully resolve $r_g$, it is unsurprising that our measurement of $\mbh$ carries a large uncertainty of about 40\% (dominated mostly by the uncertainties from the dust correction and host galaxy modeling), and that our dynamical models have a degeneracy between BH and stellar mass. In essence, our dynamical models' ability to distinguish their separate contributions is severely limited when $r_g$ is not fully resolved. This limitation was demonstrated by our test removal of the innermost component of the host galaxy MGE model, as the increase in $\mbh$ was commensurate with the decrease in stellar mass. While the total enclosed mass is well-constrained at large radii, the mass contributions from the BH and the stars within the central regions are not. This degeneracy is highlighted in Figure \ref{fig:ngc1380MGEenclosedmass}, which shows stellar mass versus radius for our dust-masked and dust-corrected host galaxy models. With each progressive increase in assumed dust extinction, the stellar mass becomes a larger fraction of the total enclosed mass. While the differences in our stellar mass profiles are minimal at radii greater than ${\sim}$100 pc, it is their differences within the innermost ${\sim}$30 pc that lead to decreases in $\mbh$ and variations in $\Upsilon_H$. Given that the total enclosed mass is tightly constrained by the well-resolved kinematics at large radii, the cuspier surface brightness models require smaller $\mbh$ values. 

For NGC 6861, we estimated $r_g$ using the best-fit values of $\mbh$ from both model E and model F. The BH mass and the enclosed stellar mass were equal at 47 pc ($0\farcs{36}$) if we adopt the stellar mass profile from model E, and at 94 pc ($0\farcs{72}$) for model F. If we instead use the velocity dispersion of $\sigma_{\star} = 389\,\mathrm{km}\,\mathrm{s^{-1}}$ from \cite{2013AJ....146...45R}, we obtain $r_g = 32$ pc ($0\farcs{25}$) when adopting the best-fit $\mbh$ from model E and 82 pc ($0\farcs{63}$) from model F. Considering that the average ALMA beam size for the NGC 6861 data is $0\farcs{28}$, these estimates suggest that $r_g$ would be resolved in NGC 6861.

Despite having observations that could in principle resolve $r_g$ if the gas disk extended to the center in NGC 6861, the lack of CO emission within the innermost ${\sim} 1\arcsec{}$ precludes a high-precision BH mass measurement, as our two dynamical models found BH mass values that differed by a factor of ${\sim}3 $ and were degenerate with $\Upsilon_H$. We attribute this large disparity to the central hole, and to the differences between our Nuker and Core-Sérsic interpolation MGE models in the dust-affected regions. These differences, especially in the slope of the surface brightness profile, led to distinct $\Upsilon_H$ values in the dynamical models. Additionally, the dearth of CO emission within the innermost $1\arcsec{}$ meant that model fits were optimized over pixels that were more sensitive to differences in the stellar mass distribution. Figure \ref{fig:mass_coresersic_nuker} shows the separate and combined enclosed mass profiles of the stars, gas, and BH for each host galaxy model used. Using these mass profiles, we determined the total enclosed mass within the central hole (which is divided between contributions from the BH and the stars, due to the absence of gas) to be $7.46 \times 10^9 \,M_{\odot}$ ($\mbh = 1.13 \times 10^9\,M_{\odot}$) when using the Nuker interpolation and $8.74 \times 10^9\,M_{\odot}$ ($\mbh = 2.87 \times 10^9\,M_{\odot})$ for the Core-Sérsic interpolation, which are both consistent with results from \cite{2017ApJ...845..170B}. Considering that the available dynamical information is restricted to radii extending beyond the hole radius, and that our best-fit values of $\mbh$ represent a minor fraction of the total dynamical mass within the hole, it is unsurprising that our two dynamical models find very different but ostensibly precise values of $\mbh$. This precision is seen in the results of our Monte Carlo simulations, and it highlights the importance of accounting for these types of systematic uncertainties when making gas-dynamical BH mass measurements in this regime.

\subsection{BH Mass Comparisons}

Although there is no prior dynamical BH mass measurement for NGC 1380, \cite{2013MNRAS.433..235P} did predict $ \mbh = 2.2^{+1.8}_{-0.9} \times 10^8$ based on velocity dispersion measurements of the globular clusters in the galaxy, which is consistent with our findings. \cite{2013AJ....146...45R} measured the BH mass in NGC 6861 through stellar-dynamical modeling and found  $\mbh = (2.0 \pm 0.2) \times 10^9 \,M_{\odot}$, which is contained within our range of estimates for $\mbh$. 

Using equations 6, 7, and 8 from \cite{2013ARAA..51..511K}, we derived predictions of $\mbh$ from the estimated total $K$-band luminosity, the estimated bulge mass, and the stellar velocity dispersion for each of the galaxies. For the $\mbh - L_{\mathrm{bul},K}$ relation, we converted the total $K$-band apparent magnitudes from 2MASS into corresponding total $K$-band luminosities using our assumed luminosity distances. For NGC 1380, we adopted the measured $R$-band $B/T = 0.359$ from \cite{2019ApJS..244...34G} to determine $L_{\mathrm{bul},K}$, while we used $B/T = 1$ for NGC 6861 adopting its classification as an elliptical galaxy. \cite{2003A&A...398...89S} studied the wavelength dependence of $B/T$ in ETGs through evolutionary synthesis modeling, and found that while $B/T$ does change substantially from the $U$ through $I$ bands, the changes  diminish at redder wavelengths. In addition, \cite{2011ApJS..197...22L} determined that the observed $B-I$, $V-I$, and $R-I$ profiles in NGC 1380 remained flat over a large radial range. Based on this information, we expect that the difference between $R$ and $K$-band $B/T$ values are relatively small, and given that there are no presently available $B/T$ measurements in the $K$-band, we use the $R$-band value as an estimate in our calculations of $L_{\mathrm{bul},K}$.  To compare with the $\mbh - M_{\mathrm{bul}}$ relation, we calculated $L_{\mathrm{bul},H}$ for each galaxy by assuming $H-K = 0.2$ mag based on SSP models and an absolute $H$-band ($K$-band) magnitude of 3.37 (3.27) for the Sun \citep{2018ApJS..236...47W}, and multiplied $L_{\mathrm{bul},H}$ by our best-fit $\Upsilon_H$ values to derive an estimate for $M_{\mathrm{bul}}$. Finally, to compare our BH mass measurements to the $M_{\mathrm{BH}}-\sigma_{\star}$ relation, we used the $\sigma_{\star}$ values of 215 km $\mathrm{s^{-1}}$ and 389 km $\mathrm{s^{-1}}$ for NGC 1380 and NGC 6861, respectively. We note that the $\mbh - L_{\mathrm{bul},K}$, $\mbh - M_{\mathrm{bul}}$, and $M_{\mathrm{BH}}-\sigma_{\star}$ relations of \cite{2013ARAA..51..511K} have intrinsic scatters of 0.28, 0.30, and 0.28 dex.

With a best-fit value of $\mbh = 1.47 \times 10^8\, M_{\odot}$ and an associated uncertainty of about 40\%, our estimate of $\mbh$ in  NGC 1380 generally agrees with predictions made by the BH-host galaxy scaling relations. For NGC 1380, we derived a total $K$-band bulge luminosity of $3.8 \times 10^{10}\, L_{\odot}$ and a bulge mass of $M_{\mathrm{bul}} = 5.0 \times 10^{10}\, M_{\odot}$ to use in the $\mbh - L_{\mathrm{bul},K}$ and  $\mbh - M_{\mathrm{bul}}$ relations. These relations predict ranges of $\mbh = (1.4-2.0) \times 10^8\, M_{\odot}$ and $ \mbh = (1.8-2.5) \times 10^8\, M_{\odot}$, respectively. The range predicted from the stellar velocity dispersion of 215 km $\mathrm{s^{-1}}$ is $(3.7-4.8) \times 10^8\, M_{\odot}$. Thus, our measurement of  $\mbh$ in NGC 1380 directly overlaps with the predictions made from the $\mbh - L_{\mathrm{bul},K}$ and $\mbh - M_{\mathrm{bul}}$ relations, and lies slightly below and outside the scatter from the $M_{\mathrm{BH}}-\sigma_{\star}$ relation.

For NGC 6861, our range of $\mbh = (1-3) \times 10^9 \,M_{\odot}$ is consistent with the previous measurement of $\mbh = (2.0 \pm 0.2) \times 10^9 \,M_{\odot}$ by \cite{2013AJ....146...45R}, and generally agrees with predictions from the BH-host galaxy scaling relations.  We derived a total $K$-band bulge luminosity of $1.2 \times 10^{11}\, L_{\odot}$, which estimates $\mbh = (0.6-0.8) \times 10^9\, M_{\odot}$. Our estimated bulge mass of $M_{\mathrm{bul}} = 2.8 \times 10^{11}\, M_{\odot}$ is higher than the value provided in \cite{2013ARAA..51..511K} of $M_{\mathrm{bul}} = 1.8 \times 10^{11}\, M_{\odot}$. Using our value, the predicted range of $\mbh$ is $(1.4-2.0) \times 10^9\, M_{\odot}$, while using the lower \cite{2013ARAA..51..511K} estimate of $M_{\mathrm{bul}}$ gives $(0.8-1.1) \times 10^9\, M_{\odot}$. The stellar velocity dispersion of 389 km $\mathrm{s^{-1}}$ in NGC 6861 is one of the highest measured in an ETG. It predicts $(5.4-11.4) \times 10^9\, M_{\odot}$ from the  $M_{\mathrm{BH}}-\sigma_{\star}$ relation, which is higher than our measured value, although our measurement is still contained within the relation's intrinsic scatter. \cite{2010ApJ...711.1316M} suggests that NGC 6861 may have had strong gravitational encounters in its past with neighboring galaxies that has elevated its central velocity dispersion, and that it could be the dominant galaxy in a galaxy subgroup that is merging based on Chandra X-ray observations. While this possibility has yet to be confirmed, our measured range of $\mbh$ and the measurement by \cite{2013AJ....146...45R} suggest that the $M_{\mathrm{BH}}-\sigma_{\star}$ relation slightly overpredicts $\mbh$ in NGC 6861. Whether this excess should be attributed to larger intrinsic scatter at the high-$\sigma_{\mathrm{\star}}$ end of the relation or physical mechanisms that have affected the growth and evolution of NGC 6861 and its BH remains unclear.

\section{Conclusion}
\label{sec:Conclusion}
We present gas-dynamical measurements of the BH masses in NGC 1380 and NGC 6861 using ALMA CO(2-1) observations at $0\farcs{21}$ and $0\farcs{28}$ resolution, respectively. We find evidence for gas disks exhibiting regular rotation in the central regions of both galaxies. For NGC 1380, a slight central increase is observed in its maximum LOS velocity, reaching approximately ${\pm}300$  $\mathrm{km}\,\mathrm{s^{-1}}$ relative to the systemic velocity of the galaxy, as expected of rotation around a central BH. In NGC 6861, the presence of a rotating gas disk with ring-like structure is observed with peak LOS velocities of ${\sim}500$ $\mathrm{km}\,\mathrm{s^{-1}}$, but a central ${\sim}1\arcsec{}$ hole in the CO distribution precludes a precise measurement of $\mbh$.

For NGC 1380, we determine $\mbh = 1.47 \times 10^8\, M_{\odot}$ with an uncertainty of about 40\% by optimizing thin disk models to the ALMA observations with four different host galaxy models. We find that our measured values of $\mbh$ are degenerate with the enclosed stellar mass, and that the uncertainties associated with the dust corrections and host galaxy models dominate the error budget. Given the slight central rise in observed LOS velocity, it is possible that higher resolution ALMA observations could provide a more confident determination of the BH mass by lifting the stellar and BH mass degeneracy. 

In the case of NGC 6861, we optimize dynamical modeling fits to the ALMA CO data using two different host galaxy models, and find that the results for $\mbh$ differ by a factor of ${\sim}3$ due to the lack of dynamical tracers within the innermost $1 \arcsec{}$, and to the structural differences in the shape of the dust-corrected surface brightness and stellar mass profiles of the host galaxy. Given the large difference between the two results, the value of $\mbh$ in NGC 6861 cannot be precisely constrained, although we find that our models suggest a plausible range of $\mbh = (1-3) \times 10^9 \,M_{\odot}$ and a lower limit of ${\sim}1 \times 10^8 \,M_{\odot}$ derived by assuming an unlikely amount of central dust extinction. This range encompasses the stellar-dynamical mass measurement of $(2.0 \pm 0.2) \times 10^9 \,M_{\odot}$ determined by \cite{2013AJ....146...45R}. 

When comparing our measurements of $\mbh$ to the BH-host galaxy scaling relations determined by \cite{2013ARAA..51..511K}, we find that our measured values of $\mbh$ for NGC 1380 and NGC 6861 are generally consistent with the $\mbh - L_{\mathrm{bul}}$, $\mbh - M_{\mathrm{bul}}$ relations, and are below the expected values predicted by the $M_{\mathrm{BH}}-\sigma_{\star}$ relation, though for NGC 6861, the measured value of $\mbh$ remains within this relation's scatter. More precise BH mass measurements on the high-mass end of these scaling relations are needed to further our understanding of the differences among them and to determine whether there is more intrinsic scatter than previously thought. 

Our work highlights a number of factors that limit gas-dynamical BH mass measurements with ALMA. Factors such as dust obscuring the stellar light, a lack of high-velocity emission within $r_g$, and the presence of a hole in the CO distribution lead to degeneracies among model parameters and large systematic uncertainties that are important to account for. A key area of improvement would be to incorporate realistic 3D radiative transfer modeling codes \citep{2013A&A...550A..74D,2015A&C.....9...20C} to recover the intrinsic stellar surface brightness of the host galaxy from NIR images. This goal is especially important for ALMA observations of dusty ETGs that do not resolve gas deep within the sphere of influence. Nevertheless, ALMA observations in this regime provide high-resolution information on circumnuclear disks in ETGs and meaningful constraints on $\mbh$. These BH mass measurements will continue to add valuable information to both local BH demographics and our understanding of galaxy evolution.

\begin{acknowledgments}
This paper makes use of the following ALMA data: ADS/JAO.ALMA\#2013.1.00229.S. ALMA is a partnership of ESO (representing its member states), NSF (USA) and NINS (Japan), together with NRC (Canada), MOST and ASIAA (Taiwan), and KASI (Republic of Korea), in cooperation with the Republic of Chile. The Joint ALMA Observatory is operated by ESO, AUI/NRAO and NAOJ. The National Radio Astronomy Observatory is a facility of the National Science Foundation operated under cooperative agreement by Associated Universities, Inc. Research at UC Irvine was supported in part by NSF grant AST-1614212. Support for this
work was provided by the NSF through awards SOSPA6-003 and SOSPADA-002 from the NRAO. Support for HST program \#15226 was provided by NASA through a grant from the Space Telescope Science Institute, which is operated by the Association of Universities for Research in Astronomy, Inc., under NASA contract NAS5-26555. Research at Texas A\&M was supported in part by NSF grant AST-1814799. LCH  was supported by the National Science Foundation of China (11721303, 11991052), China Manned Space Project (CMS-CSST-2021-A04), and the National Key R\&D Program of China (2016YFA0400702). We thank the anonymous referee for their constructive feedback.

\end{acknowledgments}

\facilities{HST (WFC3), ALMA}
\software{astropy (The Astropy Collaboration 2013, 2018), CASA (McMullin et al. 2007),  
LMFIT (Newville et al. 2016), MgeFit (Emsellem et al. 1994; Cappellari 2002), GALFIT (Peng et al. 2002), Tiny Tim (Krist \& Hook 2004), JamPy (Cappellari 2008), scikit-image (Van der Walt et al. 2014)}

\appendix 

In Table \ref{tab:NGC1380ExtraMGEs} and \ref{tab:NGC6861ExtraMGEs}, we list the components of the dust-masked and dust-corrected MGEs for NGC 1380 and NGC 6861 described in Sections \ref{sec:NGC1380HostGalaxyModels} and \ref{sec:NGC6861HostGalaxyModels}, which were used in dynamical models A, C, D, and F.

\begin{deluxetable*}{c|ccc|ccc|ccc}[htbp]
\tabletypesize{\footnotesize}
\tablecaption{NGC 1380 $H$-band MGE Parameters}
\tablewidth{0pt}
\tablehead{
\multicolumn{1}{c|}{$k$} &
\colhead{$\log_{10}$ $I_{H, k}$ ($L_{\odot}\, \mathrm{pc}^{-2}$)} &
\colhead{$\sigma_{k}^{\prime}$ (arcsec)} &
\multicolumn{1}{c|}{$q_{k}^\prime{}$} & \colhead{$\log_{10}$ $I_{H, k}$ ($L_{\odot}\, \mathrm{pc}^{-2}$)} &
\colhead{$\sigma_{k}^{\prime}$ (arcsec)} &   
\multicolumn{1}{c|}{$q_{k}^\prime{}$} & \colhead{$\log_{10}$ $I_{H, k}$ ($L_{\odot}\, \mathrm{pc}^{-2}$)} & \colhead{$\sigma_{k}^{\prime}$ (arcsec)} &\colhead{$q_{k}^\prime{}$}
\\[-1.5ex]
\multicolumn{1}{c|}{(1)} & \colhead{(2)} & \colhead{(3)} & 
\multicolumn{1}{c|}{(4)} & \colhead{(5)} & \colhead{(6)} & 
\multicolumn{1}{c|}{(7)} & \colhead{(8)} & \colhead{(9)} & \colhead{(10)}}
\startdata
  & \multicolumn{9}{c}{\bf NGC 1380}  \\
  & \multicolumn{3}{c}{$A_H=0.00$ mag} & \multicolumn{3}{c}{$A_H=0.75$ mag} & \multicolumn{3}{c}{$A_H=1.50$ mag} \\ \cline{2-4} \cline{5-7} \cline{8-10}
1 & 4.931 & 0.094  & 0.988  & 4.998  & 0.145 & 0.807 & 3.639 & 0.542 & 0.400 \\
2 & 4.968 & 0.193  & 0.761  & 5.648  & 0.052 & 0.677 & 5.669 & 0.068 & 0.745 \\
3 & 4.700 & 0.409  & 0.840  & 4.859  & 0.303 & 0.648 & 5.073 & 0.251 & 0.706 \\
4 & 4.389 & 0.879  & 0.752  & 4.628  & 0.610 & 0.802 & 4.669 & 0.595 & 0.789 \\
5 & 4.359 & 1.375  & 0.809  & 4.433  & 1.346 & 0.797 & 4.459 & 1.324 & 0.788  \\
6 & 4.000 & 3.280  & 0.608  & 4.019  & 3.214 & 0.634 & 4.032 & 3.151 & 0.642  \\
7 & 3.462 & 3.493  & 0.999  & 3.401  & 3.622 & 0.999 & 3.408 & 3.528 & 0.999   \\
8 & 3.734 & 6.119  & 0.720  & 3.734  & 6.151 & 0.702 & 3.727 & 6.223 & 0.691 \\
9 & 3.381 & 12.981 & 0.715  & 3.353  & 13.136 & 0.740 & 3.352 & 13.185 & 0.737  \\
10 & 3.068  & 15.372 & 0.400 & 3.037 & 18.912 & 0.400 & 3.050 & 18.752 & 0.400  \\
11 & 2.711  & 41.875  & 0.400 & 2.730 & 41.804 & 0.400 & 2.749 & 41.369 & 0.400 \\
12 & 2.179  & 53.560  & 0.785 & 2.401 & 49.166 & 0.642 & 2.462 & 48.099 & 0.639   \\
13 & 1.757  & 68.038  & 0.449 & 2.137 & 57.557 & 0.400 & 2.215 & 54.962 & 0.401 
\enddata
\label{tab:NGC1380ExtraMGEs}
\tablecomments{Additional NGC 1380 MGE solutions built from the HST $H$-band image. These MGEs were used to optimize dynamical models A, B, and D. These solutions have assumed central extinction values of $A_H = 0.00, 0.75$, and $1.50$ mag. The first column is the component number, the second is the central surface brightness corrected for Galactic extinction and assuming an absolute solar magnitude of $M_{{\odot},H} = 3.37$ mag \citep{2018ApJS..236...47W}, the third is the Gaussian's standard deviation along the major axis, and the fourth is the axial ratio, which was constrained to have a minimum value of 0.400 to allow for a broader range in the inclination angle during the deprojection process. Primes indicate projected quantities.}
\end{deluxetable*}

\begin{deluxetable*}{c|ccc}[htbp]
\tabletypesize{\footnotesize}
\tablecaption{NGC 6861 $H$-band MGE Parameters}
\tablewidth{0pt}
\tablehead{
\multicolumn{1}{c|}{$k$} &
\colhead{$\log_{10}$ $I_{H, k}$ ($L_{\odot}\, \mathrm{pc}^{-2}$)} &
\colhead{$\sigma_{k}^{\prime}$ (arcsec)} & \colhead{$q_{k}^\prime{}$}
\\[-1.5ex]
\multicolumn{1}{c|}{(1)} & \colhead{(2)} & \colhead{(3)} & 
\multicolumn{1}{c}{(4)}}
\startdata
  & \multicolumn{3}{c}{\bf NGC 6861}  \\
  & \multicolumn{3}{c}{$A_H=0.00$ mag (Core-Sérsic Model)}  \\ \cline{2-4} 
1 & 4.991 & 0.17  & 0.562  \\
2 & 4.183 & 0.502  & 0.567  \\
3 & 5.580 & 0.060  & 0.481   \\
4 & 4.600 & 0.343  & 0.554 \\
5 & 4.005 & 0.547  & 0.542   \\
6 & 4.097 & 1.046  & 0.623    \\
7 & 4.255 & 0.843  & 0.535    \\
8 & 4.153 & 2.254  & 0.593    \\
9 & 4.148 & 3.581  & 0.556   \\
10 & 4.174 & 1.677  & 0.501   \\
11 & 3.800 & 6.991  & 0.508 \\
12 & 3.305 & 11.455  & 0.635  \\
13 & 1.368 & 21.274  & 0.988  \\
14 & 2.619 & 24.889 & 0.999
\enddata
\label{tab:NGC6861ExtraMGEs}
\tablecomments{The NGC 6861 $A_H=0.00$ mag (Core-Sérsic Model) MGE solution built from the HST $H$-band image. This MGE solution was used to optimize dynamical model F. The first column is the component number, the second is the central surface brightness corrected for Galactic extinction and assuming an absolute solar magnitude of $M_{{\odot},H} = 3.37$ mag \citep{2018ApJS..236...47W}, the third is the Gaussian standard deviation along the major axis, and the fourth is the axial ratio, which was constrained to have a minimum value of 0.400 to allow for a broader range in the inclination angle during the deprojection process. Primes indicate projected quantities.}
\end{deluxetable*}


\vspace{5mm}
% \facilities{HST(STIS), Swift(XRT and UVOT), AAVSO, CTIO:1.3m,


%\software{astropy %\citep{2013A&A...558A..33A},LMFIT \citep{}}



\bibliography{paperbibliography}{}
\bibliographystyle{aasjournal}



\end{document}